\documentclass[twocolumn]{aastex63}
\usepackage{lineno}
\defcitealias{Tautvaisiene20}{Paper~I}

\newcommand\noallfield{848}
\newcommand\noatmospheric{740}
\newcommand\nobinary{28}

\received{2022 January 7}
\revised{2022 January 27}
\accepted{2022 January 29}
\submitjournal{ApJS}

\shorttitle{Chemical composition of bright stars}
\shortauthors{Tautvai\v{s}ien\.{e} et al.}

\begin{document}

\title{CHEMICAL COMPOSITION OF BRIGHT STARS IN THE NORTHERN HEMISPHERE: STAR--PLANET CONNECTION}

\correspondingauthor{Gra\v{z}ina Tautvai\v{s}ien\.{e}}
\email{grazina.tautvaisiene@tfai.vu.lt}

\author{G. Tautvai\v{s}ien\.{e}}
\affiliation{Astronomical Observatory, Institute of Theoretical Physics and Astronomy, 
Vilnius University, \\ Sauletekio av. 3, 10257 Vilnius, Lithuania}

\author{\v{S}. Mikolaitis}
\affiliation{Astronomical Observatory, Institute of Theoretical Physics and Astronomy, 
Vilnius University, \\ Sauletekio av. 3, 10257 Vilnius, Lithuania}

\author{A. Drazdauskas}
\affiliation{Astronomical Observatory, Institute of Theoretical Physics and Astronomy, 
Vilnius University, \\ Sauletekio av. 3, 10257 Vilnius, Lithuania}

\author{E. Stonkut\.{e}}
\affiliation{Astronomical Observatory, Institute of Theoretical Physics and Astronomy, 
Vilnius University, \\ Sauletekio av. 3, 10257 Vilnius, Lithuania}

\author{R. Minkevi\v{c}i\={u}t\.{e}}
\affiliation{Astronomical Observatory, Institute of Theoretical Physics and Astronomy, 
Vilnius University, \\ Sauletekio av. 3, 10257 Vilnius, Lithuania}

\author{E. Pak\v{s}tien\.{e}}
\affiliation{Astronomical Observatory, Institute of Theoretical Physics and Astronomy, 
Vilnius University, \\ Sauletekio av. 3, 10257 Vilnius, Lithuania}

\author{H. Kjeldsen}
\affiliation{Astronomical Observatory, Institute of Theoretical Physics and Astronomy, 
Vilnius University, \\ Sauletekio av. 3, 10257 Vilnius, Lithuania}
\affiliation{Stellar Astrophysics Centre, Department of Physics and Astronomy, Aarhus University, \\ Ny Munkegade 120, DK-8000 Aarhus C, Denmark}

\author{K. Brogaard}
\affiliation{Astronomical Observatory, Institute of Theoretical Physics and Astronomy, 
Vilnius University, \\ Sauletekio av. 3, 10257 Vilnius, Lithuania}
\affiliation{Stellar Astrophysics Centre, Department of Physics and Astronomy, Aarhus University, \\ Ny Munkegade 120, DK-8000 Aarhus C, Denmark}

\author{Y. Chorniy}
\affiliation{Astronomical Observatory, Institute of Theoretical Physics and Astronomy, 
Vilnius University, \\ Sauletekio av. 3, 10257 Vilnius, Lithuania}

\author{C. von Essen}
\affiliation{Astronomical Observatory, Institute of Theoretical Physics and Astronomy, 
Vilnius University, \\ Sauletekio av. 3, 10257 Vilnius, Lithuania}
\affiliation{Stellar Astrophysics Centre, Department of Physics and Astronomy, Aarhus University, \\ Ny Munkegade 120, DK-8000 Aarhus C, Denmark}

\author{F. Grundahl}
\affiliation{Astronomical Observatory, Institute of Theoretical Physics and Astronomy, 
Vilnius University, \\ Sauletekio av. 3, 10257 Vilnius, Lithuania}
\affiliation{Stellar Astrophysics Centre, Department of Physics and Astronomy, Aarhus University, \\ Ny Munkegade 120, DK-8000 Aarhus C, Denmark}

\author{M. Ambrosch}
\affiliation{Astronomical Observatory, Institute of Theoretical Physics and Astronomy, 
Vilnius University, \\ Sauletekio av. 3, 10257 Vilnius, Lithuania}

\author{V. Bagdonas}
\affiliation{Astronomical Observatory, Institute of Theoretical Physics and Astronomy, 
Vilnius University, \\ Sauletekio av. 3, 10257 Vilnius, Lithuania}

\author{A. Sharma}
\affiliation{Astronomical Observatory, Institute of Theoretical Physics and Astronomy, 
Vilnius University, \\ Sauletekio av. 3, 10257 Vilnius, Lithuania}

\author{C. Viscasillas V\'azquez}
\affiliation{Astronomical Observatory, Institute of Theoretical Physics and Astronomy, 
Vilnius University, \\ Sauletekio av. 3, 10257 Vilnius, Lithuania}


 
\begin{abstract}

In fulfilling the aims of the planetary and asteroseismic research missions, such as that of the NASA Transiting Exoplanet Survey Satellite (TESS) space telescope, accurate stellar atmospheric parameters and a detailed chemical composition are required as input. 
We have observed high-resolution spectra for all \noallfield ~bright ($V<8$~mag) stars that are cooler than F5 spectral class in the area up to 12~deg surrounding the northern TESS continuous viewing zone and uniformly determined the main atmospheric parameters, ages, orbital parameters, velocity components, and precise abundances of up to 24 chemical species 
(C(C$_2$), N(CN), [\ion{O}{1}], \ion{Na}{1}, \ion{Mg}{1}, \ion{Al}{1}, \ion{Si}{1}, \ion{Si}{2}, \ion{Ca}{1}, \ion{Ca}{2}, \ion{Sc}{1}, \ion{Sc}{2}, \ion{Ti}{1}, \ion{Ti}{2}, \ion{V}{1}, \ion{Cr}{1}, \ion{Cr}{2}, \ion{Mn}{1}, \ion{Fe}{1}, \ion{Fe}{2}, \ion{Co}{1}, \ion{Ni}{1}, \ion{Cu}{1}, and \ion{Zn}{1}) for \noatmospheric ~slowly rotating stars. 
The analysis of 25 planet-hosting stars in our sample drove us to the following conclusions: the  dwarf  stars  hosting  high-mass  planets  are more  metal  rich  than  those  with  low-mass  planets. We find slightly negative C/O and Mg/Si slopes toward the stars with high-mass planets. All the low-mass planet hosts in our sample show positive $\Delta{\rm [El/Fe]}$ versus condensation temperature slopes, in particular, the star with the large number of various planets. The high-mass planet hosts have a diversity of slopes, but in more metal rich, older, and cooler stars, the positive elemental abundance slopes are more common.

\end{abstract}

\keywords{High resolution spectroscopy; Catalogs; Chemical abundances.}


\section{Introduction} 
\label{sec:intro}


The NASA Transiting Exoplanet Survey Satellite (TESS) is an ongoing space mission with the primary goal of searching for planets in systems of bright and nearby stars as well as providing precise asteroseismic information \citep{Ricker2015}. 
The first work of this series (\citealt[hereafter Paper~I]{Tautvaisiene20}) was dedicated to observations of bright stars in the TESS northern continuous viewing zone (CVZ). We observed high-resolution spectra for all stars up to $V<8$~mag and cooler than F5 spectral type and determined the main atmospheric parameters, ages, kinematic parameters, and abundances of up to 24 chemical elements for 277 slowly rotating stars. In the current work, we extend the homogeneous analysis by observing all \noallfield ~stars located around the TESS CVZ up to 12~deg and increasing the number of bright stars with determined parameters and chemical composition by \noatmospheric ~and the total sample of stars up to 1017. Similar observations and analyses were previously done by us for bright dwarf stars in two preliminary ESA PLATO space mission fields (\citealt{Mikolaitis2018, Mikolaitis2019, Stonkute2020}). As the number of bright stars with confirmed planets in the covered sky areas has already increased up to 25, we decided to address several questions about the star-planet connection that are currently under discussion in the literature: 
the stellar chemical composition and planet mass relation (e.g. \citealt{Santos2017a, Suarez2017,Suarez18, Bedell2018, Hinkel2018, Adibekyan2019, Cridland2019,  Bashi2022, DelgadoMena2021,Kolecki2021, Mishenina2021}); and elemental abundance versus condensation temperature ($T_{\rm c}$) relations in planet-hosting stars (e.g. \citealt{daSilva2015, Bedell2018, Liu2020,Cowley2021, Mishenina2021}). The large sample of homogeneously investigated comparison stars in our study allows to take into account spatial and temporal factors as well as other specificities of the Galactic and stellar evolution.   
 
\section{Observations and method of analysis} \label{sec:obs-methods}

\begin{figure}
\epsscale{1.17}
\plotone{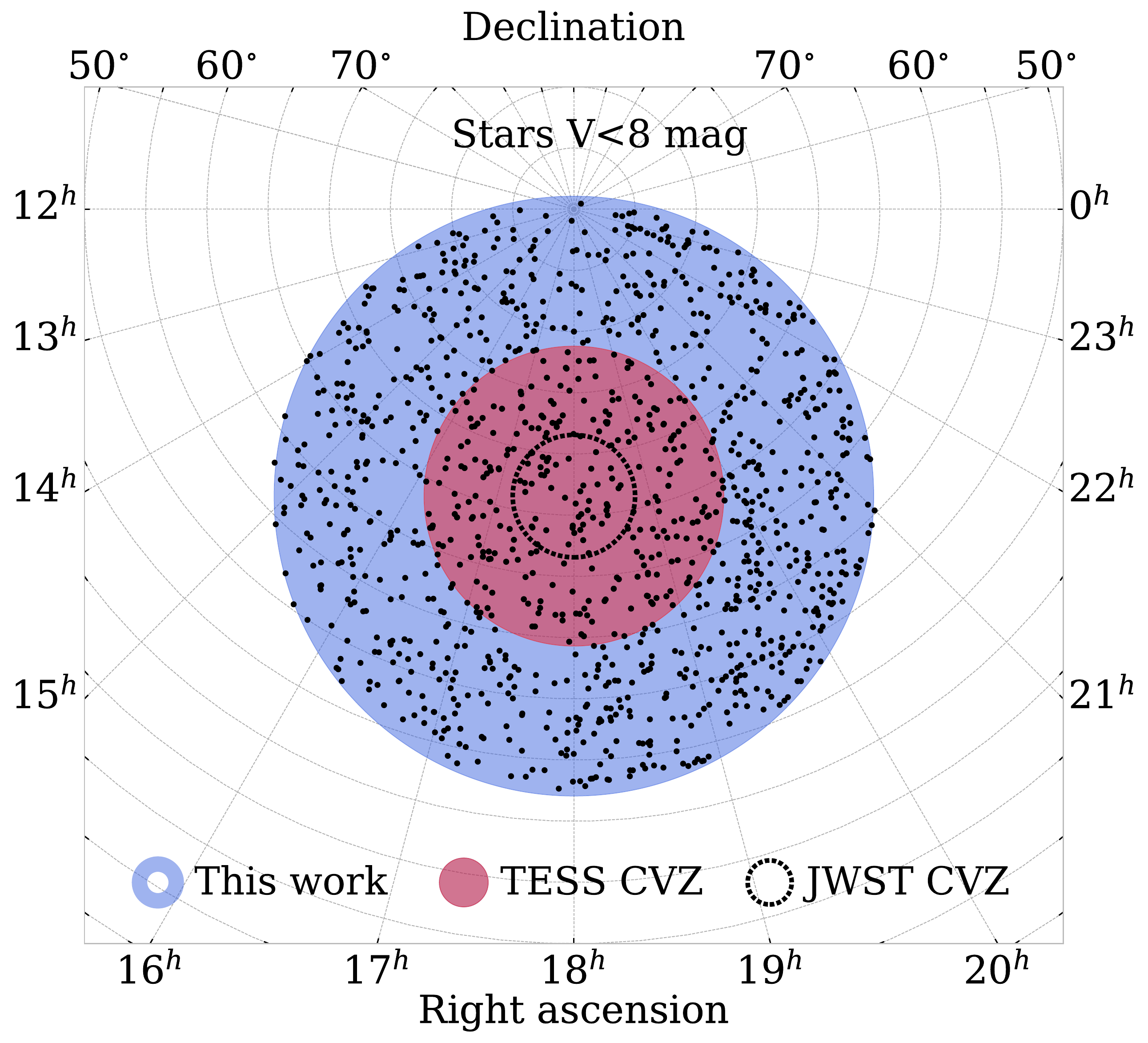}
\caption{Positions (R.A. and decl. in hours and degrees) of the stars (black dots) observed in this work (blue shadowed area) and in \citetalias{Tautvaisiene20} (the pink shadowed area) in the TESS CVZ. The JWST CVZ is indicated as well.
\label{fig:CVZ}}
\end{figure}

\subsection{Observations and Target Selection}\label{subsec:targets}

Observations were carried out with the 1.65~m telescope at the Moletai Astronomical Observatory of Vilnius University in Lithuania, which is equipped with the high-resolution Vilnius University Echelle Spectrograph (VUES; \citealt{Jurgenson2016}). This spectrograph has a wavelength coverage from 400 to 900~nm in $R\sim$36,000, $\sim$51,000, and $\sim$68,000 resolution modes. For our work, we used the $\sim 68,000$ mode for the M spectral type stars and the $\sim 36,000$ mode for other objects. Exposure times varied between 900 and 2400~s and signat-to-noice ratios (S/Ns) varied between 75 and 200 with the median value at 96, depending on stellar magnitudes.  The VUES data reduction was accomplished on site using the automated pipeline described by \citet{Jurgenson2016}.

Like in \citetalias{Tautvaisiene20}, we selected all bright ($V<8$~mag) F5 and cooler than $T_{\rm eff}$~$<$~6500~K (corresponding to approximately $(B-V)>0.39$~mag) stars in the area surrounding previously observed TESS northern CVZ up to $12^{\circ}$ around the northern ecliptic pole. 
In this way, we found \noallfield ~stars in the selected field that met these criteria (see Figure~\ref{fig:CVZ}), and we have observed all of them during the period of 2019--2021.

\subsection{Radial Velocity Determination and Identification of
Double-line Binaries and Fast-rotating Stars} \label{subsec:kinematic}

\begin{figure}
\epsscale{1.17}
\plotone{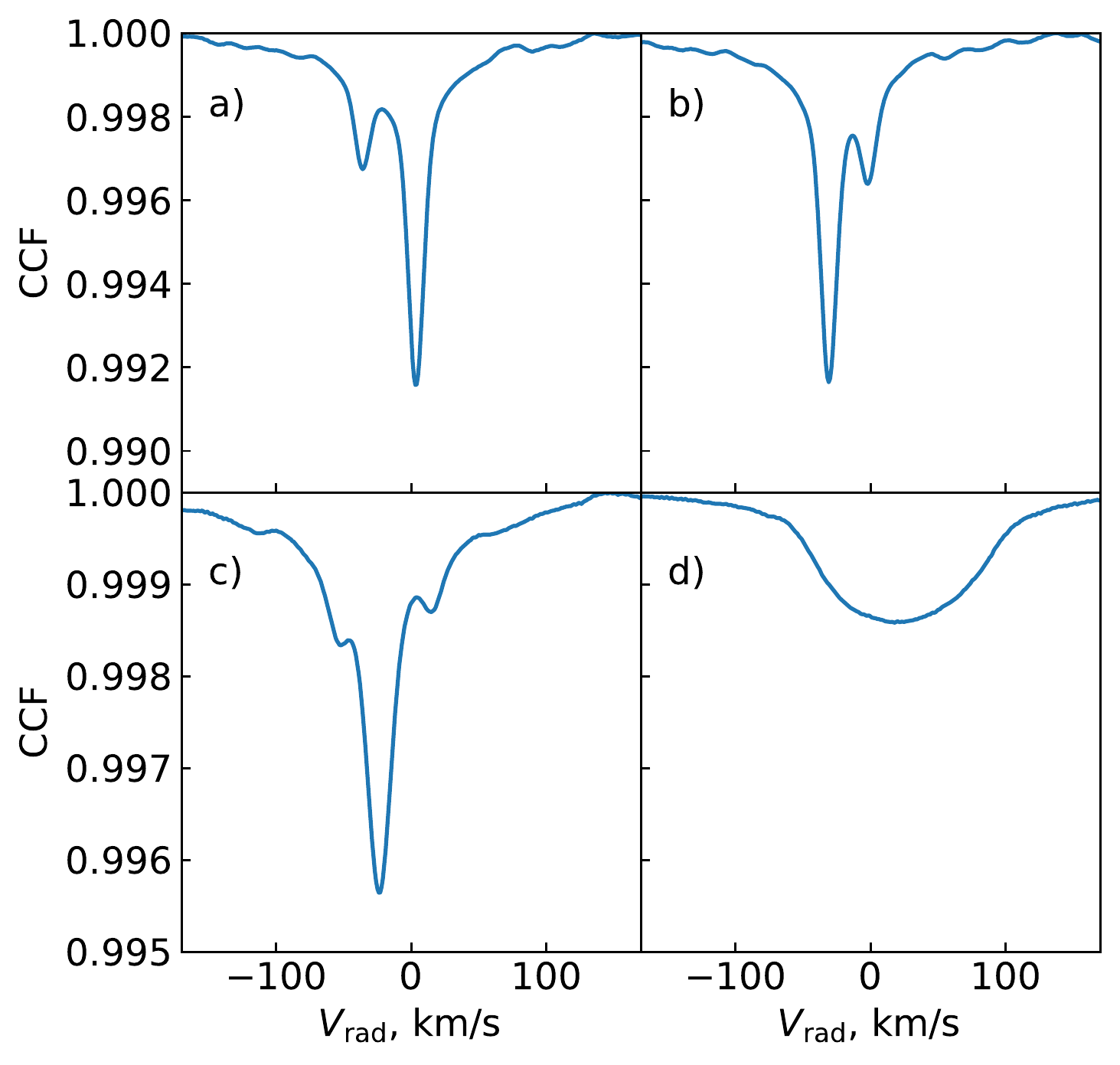}
\caption{Examples of CCFs produced to
calculate the radial velocities and detection of double-line binary stars: (a) the double-line spectroscopic binary HD~152274 observed on 2018 September 19, (b) the same HD~152274 star observed on 2019 June 28, (c)	HD~160780 showing three profiles, and (d) the fast-rotating system HD~193341. } 
\label{fig:ccf}
\end{figure}

For an initial spectral analysis, we used 
the standard cross-correlation function (CCF) method to obtain spectroscopic radial velocity values. 
The CCF revealed 27~double-line and one triple-line stellar systems. 
All \nobinary~stars with double or multiple-line features are recognized as binary systems from proper motion anomaly in Gaia and Hipparcos data (\citealt{Kervella2019}).
Of the \nobinary, 13 are already labeled as spectroscopic binaries, e.g., in Geneva--Copenhagen survey (\citealt{Nordstrom2004}) or in the SIMBAD database (\citealt{Wenger2000}). The remaining 15 stars are newly detected spectroscopic binaries.
Figure~\ref{fig:ccf} shows the CCF examples of our study.
The CCF also revealed 33~fast-rotating stars ($V_{rot}\ge 20$~km\,s$^{-1}$) with strongly broadened and diminished lines that prevented us from analyzing them.
From the subsequent analysis, we also excluded the 47 coolest (M-type) stars with severe line-blending. We postponed a further investigation of these stars. This investigation requires different methods of analysis and additional photometric and spectral observations.

Thus, of the observed \noallfield ~stars, we fully characterized a sample of \noatmospheric ~stars.

\subsection{Stellar Atmospheric Parameters and Chemical Composition} \label{subsec:parameters}

In order to determine the main stellar atmospheric parameters (effective temperature, $T_{\rm eff}$; surface gravity, ${\rm log}~g$; microturbulence velocity, $v_{\rm t}$; and metallicity ${\rm [Fe/H]}$), we adopted the  classical method of the equivalent widths of atomic neutral and ionized iron lines. 
We used a combination  of the DAOSPEC (\citealt{Stetson2008}) and MOOG (\citealt{Sneden1973}) codes the same way as the Vilnius node used in the Gaia-ESO Survey (see \citealt{Smiljanic2014} and \citealt{Mikolaitis2018}).

Detailed abundances of 24 chemical species were determined applying a spectral synthesis method with the TURBOSPECTRUM code (\citealt{Alvarez1998}). The spectral analysis was done using a grid of MARCS stellar atmosphere models \citep{Gustafsson2008} and the solar abundances by \citet{Grevesse07}. 
Atomic lines were selected from the Gaia-ESO line-list by \citet{Heiter2015}. We have also used the molecular line lists for $\rm{C}_{2}$~\citep[][]{Brooke2013,Ram2014}, CN~\citep[][]{Sneden2014}, CH~\citep[][]{Masseron2014}, SiH~\citep[][]{Kurucz1993}, FeH~\citep[][]{Dulick2003}, CaH~(B. Plez 2022, private communication), and OH, MgH, NH~(T. Masseron 2022, private communication). 
For the carbon abundance determination, we used two regions: the ${\rm C}_2$ Swan (1, 0) band head at 5135~{\AA} and the ${\rm C}_2$ Swan (0, 1) band head at 5635~{\AA}. For the nitrogen abundance determination, we used $\mathrm{^{12}C^{14}N}$ molecular lines in the regions 6470--6485 and 7980--8005~\AA. 
The oxygen abundance was determined from the forbidden [O\,{\sc i}] line at 6300~{\AA}. These elements require a more detailed analysis, as they are bound by the molecular equilibrium. 
First, we performed several iterations until the determinations of carbon and oxygen abundances converged. After this, we used both carbon and oxygen values to determine the abundance of nitrogen.

\begin{figure*}
\epsscale{1.17}
\plotone{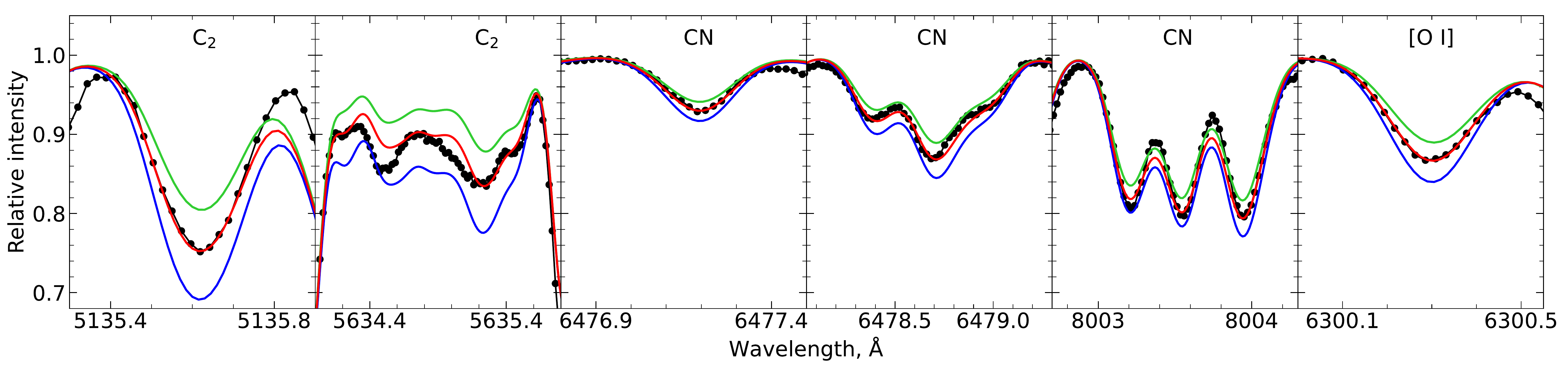}
\caption{Examples of synthetic spectrum fits to the C$_2$ band heads at 5135 and 5635~\AA, to the CN bands at 6478 and 8003~\AA, and to the forbidden [O\,{\sc i}] line at 6300~\AA. The observed spectra are shown as the solid black lines with dots. The solid red lines represent the best fits, while the solid blue and green lines represent a change in abundance by $\pm0.1$~dex of the corresponding elements.}
\label{fig:fits}
\end{figure*}

In Figure~\ref{fig:fits} we show examples of the observed and modeled C$_2$, CN, and [O\,{\sc i}] line fits. For more details of the chemical composition analysis, we refer to \citetalias{Tautvaisiene20} and other recent studies (\citealt{Mikolaitis2019}; \citealt{Stonkute2020}).

\subsection{Stellar Ages}
\label{subsec:ages}

In order to calculate stellar ages, we used the code UniDAM (the unified tool to estimate distances, ages, and masses) by \citet{Mints2017, Mints18}. The code uses a Bayesian approach and the PARSEC isochrones \citep{Bressan2012}. As an input, we used the stellar atmospheric parameters determined in this work together with the \textit{J}, \textit{H}, and \textit{K} magnitudes from the Two Micron All-Sky Survey (2MASS, \citealt{Skrutskie2006}) and the $W1$ and $W2$ magnitudes from AllWISE \citep{Cutri14}. 

We cross-matched our sample of 740 stars with the 2MASS and AllWISE catalogs and obtained 715 objects with entries in both databases. After calculating the ages, we discarded 137 stars for which the reported flags meant either an unreliable photometry, that the result was off the model grid, or just an unreliable determination (see Section 6.1 in \citealt{Mints2017} for more interpretations). Finally, we were left with 578 stars for which we report the derived ages in this work. However, one should take into account that uniDAM assumes a scaled solar abundance pattern, and this can introduce a bias in the age estimates when this assumption is wrong.

\subsection{Kinematic Properties}
\label{subsec:kinematics}

The main kinematic parameters ($R\mathrm{_{mean}}$, $z\mathrm{_{max}}$, $e$, \textit{U, V, {\rm and} W}) for the stars were calculated using the Python-based package for galactic-dynamics calculations \textit{galpy}\footnote{http://github.com/jobovy/galpy} by \citet{Bovy15}. We used two main sources for the input data: the distances were taken from \citet{Bailer21}, whereas proper motions, coordinates, and radial velocities were taken from the Gaia data release 3 (EDR3) catalog \citep{Gaia16, Gaia21, Lindegren21, Seabroke21}. The radial velocities of 686 stars were taken from the Gaia EDR3, and those of the remaining 54 were gathered either from the  SIMBAD database or from our own determinations. Several stars did not have proper motion values in the EDR3. In that case, we used values from the online SIMBAD database.

The \textit{galpy} was set to integrate orbits for 5~Gyr. Observational errors were estimated using 1000~Monte Carlo calculations according to the errors in the input parameters. The position and movements of the Sun are those from (\citealt{Bovy2012}; $R_{\rm gc\odot}=8$~kpc and $V_{\odot}=220$~km\,s$^{-1}$), the distance from the Galactic plane $z_{\odot}=0.02$~kpc \citep{Joshi07}, and the LSR from (\citealt{Schonrich10}; \textit{U, V, W} = 11.1, 12.24, 7.25~km\,s$^{-1}$).

\subsection{Errors on Atmospheric Parameters and Abundances}
\label{sec:erroratmospheres}

\begin{table*}
\caption{Median Effects on the Derived Abundances Resulting from the Atmospheric Parameter Uncertainties for the Sample Stars.
}
\label{tab:sensitivity}
      \[
         \begin{tabular}{lcrrcrrcc}
            \hline\hline
            \noalign{\smallskip}
	    El & 
	    ${ \Delta T_{\rm eff} }$ (K)& 
            ${ \Delta \log g }$ & 
            $\Delta {\rm [Fe/H]}$ & 
            ${ \Delta v_{\rm t} }$ (km\,s$^{-1}$) 
            & $\sigma_{\rm{scat}}$$^{\rm a}$  
     	     & $N_{\rm max}$$^{\rm b}$ 
     	     & $ \sigma_{\rm total\left[\frac{El}{H}\right]}$$^{\rm c} $  
	     & $ \sigma_{\rm all\left[\frac{El}{H}\right]}$$^{\rm d} $  
	     \\ 
            \noalign{\smallskip}
            \hline
            \noalign{\smallskip}

C(C$_{2}$)	&	0.00	&	0.03	&	0.01	&	0.01	&	0.02	&	2	&	0.03	&	0.04	\\
N(CN)	&	0.06	&	0.05	&	0.01	&	0.01	&	0.05	&	7	&	0.08	&	0.09	\\
O([O\,{\sc i}])	&	0.01	&	0.08	&	0.07	&	0.00	&	0.06	&	1	&	0.11	&	0.12	\\
\ion{Na}{1}	&	0.01	&	0.03	&	0.03	&	0.02	&	0.06	&	4	&	0.05	&	0.08	\\
\ion{Mg}{1}	&	0.02	&	0.07	&	0.03	&	0.03	&	0.07	&	5	&	0.08	&	0.11	\\
\ion{Al}{1}	&	0.02	&	0.03	&	0.03	&	0.03	&	0.08	&	5	&	0.06	&	0.10	\\
\ion{Si}{1}	&	0.01	&	0.02	&	0.02	&	0.02	&	0.05	&	14	&	0.04	&	0.06	\\
\ion{Si}{2}	&	0.02	&	0.06	&	0.02	&	0.02	&	0.08	&	7	&	0.07	&	0.11	\\
\ion{Ca}{1}	&	0.03	&	0.07	&	0.02	&	0.02	&	0.07	&	28	&	0.08	&	0.11	\\
\ion{Ca}{2}	&	0.03	&	0.06	&	0.02	&	0.04	&	0.06	&	3	&	0.08	&	0.10	\\
\ion{Sc}{1}	&	0.05	&	0.03	&	0.03	&	0.03	&	0.09	&	7	&	0.07	&	0.12	\\
\ion{Sc}{2}	&	0.02	&	0.07	&	0.02	&	0.04	&	0.05	&	7	&	0.09	&	0.10	\\
\ion{Ti}{1}	&	0.04	&	0.04	&	0.02	&	0.02	&	0.06	&	75	&	0.06	&	0.09	\\
\ion{Ti}{2}	&	0.01	&	0.07	&	0.04	&	0.04	&	0.05	&	19	&	0.09	&	0.10	\\
\ion{V}{1}	&	0.03	&	0.01	&	0.02	&	0.04	&	0.06	&	8	&	0.05	&	0.08	\\
\ion{Cr}{1}	&	0.02	&	0.03	&	0.02	&	0.03	&	0.06	&	7	&	0.05	&	0.08	\\
\ion{Cr}{2}	&	0.02	&	0.07	&	0.02	&	0.04	&	0.05	&	2	&	0.09	&	0.10	\\
\ion{Mn}{1}	&	0.03	&	0.03	&	0.02	&	0.04	&	0.06	&	14	&	0.06	&	0.09	\\
\ion{Fe}{1}	&	0.01	&	0.03	&	0.02	&	0.03	&	0.04	&	137	&	0.05	&	0.06	\\
\ion{Fe}{2}	&	0.01	&	0.08	&	0.03	&	0.04	&	0.07	&	11	&	0.09	&	0.12	\\
\ion{Co}{1}	&	0.01	&	0.01	&	0.01	&	0.02	&	0.07	&	7	&	0.03	&	0.07	\\
\ion{Ni}{1}	&	0.01	&	0.02	&	0.01	&	0.04	&	0.05	&	30	&	0.05	&	0.07	\\
\ion{Cu}{1}	&	0.03	&	0.02	&	0.02	&	0.02	&	0.07	&	6	&	0.05	&	0.08	\\
\ion{Zn}{1}	&	0.02	&	0.02	&	0.03	&	0.03	&	0.10	&	3	&	0.05	&	0.12	\\

\hline
         \end{tabular} 
      \]

Notes.\\
$^{\rm a} \sigma_{\rm{scat}}$ stands for the median line-to-line scatter.\\ 
$^{\rm b}{\rm N}_{max}$ presents the number of lines investigated.\\
$^{\rm c}\sigma_{\rm total([El/H])} $ stands for the median of the quadratic sum  of all four effects on [El/H].\\
$^{\rm d}\sigma_{\rm all([El/H])} $ is a median of the combined effect of $ \sigma_{\rm total([El/H])} $ and the line-to-line scatter $\sigma_{\rm{scat}}$. 

   \end{table*}

The errors on the atmospheric parameters and abundances were estimated for every star. The procedure is described in \citetalias{Tautvaisiene20} and \citet{Mikolaitis2019}. Here we briefly recall the main information.

Tests for estimating the impact of the S/N on the error budget in our atmospheric parameter determination and chemical abundance measurements have been made in \citetalias{Tautvaisiene20} for giants and in \citet{Mikolaitis2019} for dwarf stars. We used 300 generated spectra of a star for S/N values of 25, 50, and 75 to determine the atmospheric parameter and measure abundances in order to estimate their sensitivity to the quality of the spectrum. 

Evaluation of the line-to-line scatter is a way to estimate random errors if the number of lines is large enough. A median of the standard deviation for a given element, $\sigma_{\rm{scat}}^{*}$, is presented in the sixth column of Table~\ref{tab:sensitivity}.

The uncertainties of the main atmospheric parameters were propagated into the errors of chemical abundances. The median errors of this type over the stellar sample are provided in Table~\ref{tab:sensitivity}.                         

The final error for every element for every star that is given in machine-readable Table~\ref{tab:CDS} is a quadratic sum of the effects due to the uncertainty in four atmospheric parameters and the abundance scatter given by the lines.

In this paper, we use a classical local thermodynamic equilibrium (LTE) approach for all lines of the studied elements. In the metallicity range of our sample stars,  non-LTE effects should be small (see Section~3.6 of \citealt{Mikolaitis2019} and references therein).
As we focus on [C/O] and [Mg/Si] later in the paper, it is worthwhile mentioning that C$_2$ bands that were used to determine the carbon abundance, are  not  sensitive  to  non-LTE  deviations  (\citealt{Clegg1981, Gustafsson1999}). The 6300.3~\AA\ oxygen forbidden line is known to be unaffected by non-LTE and shows little sensitivity to 3D effects (\citealt{Asplund2004, Pereira2009}). This line forms nearly in LTE and is only weakly sensitive to convection, its formation is similar in 3D radiation hydrodynamic and 3D magnetoradition-hydrodynamical solar models (\citealt{Bergemann21}).
Threfore, possible non-LTE effects on [C/O] should be very small.
It was shown by \citet{Adibekyan2017} that the non-LTE effect on the [Mg/H] ratio is expected to be from $-0.01$ to ~0.03~dex, and for [Si/H] from $-0.004$ to ~0.012~dex. This leads to a possible non-LTE correction for [Mg/Si] of 0.0 to 0.03~dex in a metallicity regime similar to that of our sample. 
For manganese and copper, we have accounted for a hyperfine splitting as described in \citet{Mikolaitis2019}.

Since the abundances of C, N, and O are bound together by the molecular equilibrium in the stellar atmospheres, in \citetalias{Tautvaisiene20} we investigated how an error in one of them typically influences the abundance determination of another. We determined that $\Delta$[O/H] = 0.10 causes $\Delta$[C/H] = 0.02 and $\Delta$[N/H] = 0.04, and $\Delta$[C/H] = 0.10 causes $\Delta$[N/H] = $-0.11$ and $\Delta$[O/H] = 0.02, while $\Delta$[N/H] = 0.10 has no effect on either the carbon or the oxygen abundances.

\section{Stellar parameters}\label{sec:results}

The determined stellar parameters are presented in Table~\ref{tab:CDS} (which is available in its entirety in machine-readable form) and are discussed in the following sections.

\subsection{Stellar Ages, Kinematic Properties, and Dependence on Galactic Subcomponents}
\label{sec:kinematicparameters}

The ages of our sample stars range from about 0.2 to 12~Gyr; the majority are close to Solar, about 5~Gyr.
The age values and uncertainties are presented in Table~\ref{tab:CDS}. 

The $U$, $V$, and $W$ velocities, distances, $R_{\rm mean}$, $z_{\rm max}$, and orbital eccentricities, \textit{e}, with the corresponding errors are presented in Table~\ref{tab:CDS}.

It is widely accepted that Galactic subcomponents such as thin and thick disks differ in a number of parameters. There are two widely used methods to separate them: kinematical (e.g. \citealt{Bensby2003, Bensby2005, Bensby2014}) and chemical (e.g. \citealt{Adibekyan2012, Recio2014}).

The method introduced by \citet{Bensby2003, Bensby2014} employs the thick-disk (TD) to thin-disk (D) probability ratios. Stars with TD/D~$>$~2 are potential thick-disk stars, stars with TD/D~$<0.5$ potentially belong to the thin disk, and stars with 0.5~$<$~TD/D~$<2.0$ are called "in-between stars". We provide this TD/D value in Table~\ref{tab:CDS} as well.

The chemical separation method can employ [\ion{Mg}{1}/\ion{Fe}{1}]  (\citealt{Adibekyan2012}; \citealt{Mikolaitis2014}), [\ion{Ti}{1}/\ion{Fe}{1}] (\citealt{Bensby2014}), or [$\alpha$/\ion{Fe}{1}] (\citealt{Recio2014}) abundance ratios. Recently, \citet{Lagarde2021} have proposed the age-chemo-kinematics approach, which we applied in this work as well.  We used [\ion{Mg}{1}/\ion{Fe}{1}] and [$\alpha$/\ion{Fe}{1}] versus [\ion{Fe}{1}/H] to separate the low-$\alpha$ from high-$\alpha$ stars that potentially belong to  the thin or thick disks, respectively. The values of [$\alpha$/\ion{Fe}{1}], which are averages of \ion{Mg}{1}, $<$\ion{Si}{1}, \ion{Si}{2}$>$, $<$\ion{Ca}{1}, \ion{Ca}{2}$>$, $<$\ion{Ti}{1}, \ion{Ti}{2}$>$, and standard errors of the mean are presented in columns 77 and 78 of Table~\ref{tab:CDS}. 

Based on chemical signatures, we separated our sample of stars into 601 thin-disk, 138 thick-disk, and one high-$\alpha$ halo star. 

Recently, studies have emerged that showed that the thick-disk stars can be even further divided into separate populations of metal-rich and metal-poor stars. The exact nature of the high-$\alpha$ metal-rich (h$\alpha$mr) stars is still debated, but few explanations have been proposed. One of them \citep{Sharma2021, Zhang2021} is that these stars were born somewhere at the edge of the Bulge, at around 3-5 kpc from Galactic center, and later migrated outward. \citet{Zhang2021} discussed that the radial migration induced by  the  coupling  between  the  bar  and  the  spiral  arms could bring its stars from the birth sites of 3-5 kpc to their current locations. 

When analyzing the APOKASC sample of stars, \citet{Lagarde2021} found two density peaks in the [alpha/Fe] versus [Fe/H] plane for thick-disk stars. They determined that the kinematics of the h$\alpha$mr thick-disk population seems  to  follow  that  of  the  thin-disk  population  more  closely than that of the h$\alpha$mp thick-disk population and came to a similar conclusion as \citet{Sharma2021} and \citet{Zhang2021} that these properties might suggest a different origin and history for these stars by migration from the inner disk or as a transition region between the old thick disk and the young thin disk.

The equations for separating the thick-disk components were derived in the work by \citet{Lagarde2021}. These equations somewhat depend on the spectroscopic survey, but they worked well in our case, and 
the thick and thin disks separate quite nicely.  We use the chemical separation in our further discussion.  As the analysis of thick-disk subpopulations is not the goal of this paper, we treated the thick disk as a single population. However, in the last column of table~\ref{tab:CDS}, we identify stars that according to our analysis could be attributed to the h$\alpha$mr component as well.

\subsection{Atmospheric Parameters and Elemental Abundances}

Our sample of \noatmospheric~slowly rotating stars has 
temperatures, $T_{\rm eff}$ between 3977 and 6414~K with a mean of 4703~K. The metallicities [Fe/H] are from $-2.25$ to 0.15~dex with a mean at $-0.34$~dex, and the surface gravity log\,$g$ is from 0.51 to 3.5 with a mean of 2.4 for giants and from 3.6 to 4.48 with a mean of 4.0 for dwarfs. 

Atmospheric parameters are listed in the Table~\ref{tab:CDS} together with elemental abundances (C(C$_2$), N(CN), [\ion{O}{1}], \ion{Na}{1}, \ion{Mg}{1}, \ion{Al}{1}, \ion{Si}{1}, \ion{Si}{2}, \ion{Ca}{1}, \ion{Ca}{2}, \ion{Sc}{1}, \ion{Sc}{2}, \ion{Ti}{1}, \ion{Ti}{2}, \ion{V}{1}, \ion{Cr}{1}, \ion{Cr}{2}, \ion{Mn}{1}, \ion{Fe}{1}, \ion{Fe}{2}, \ion{Co}{1}, \ion{Ni}{1}, \ion{Cu}{1}, and  \ion{Zn}{1}) relative to the Sun and their uncertainties for the \noatmospheric ~stars investigated in the 
present study. The abundances are presented in [Element/H] form. 

\begin{figure}

 \includegraphics[width=\columnwidth]{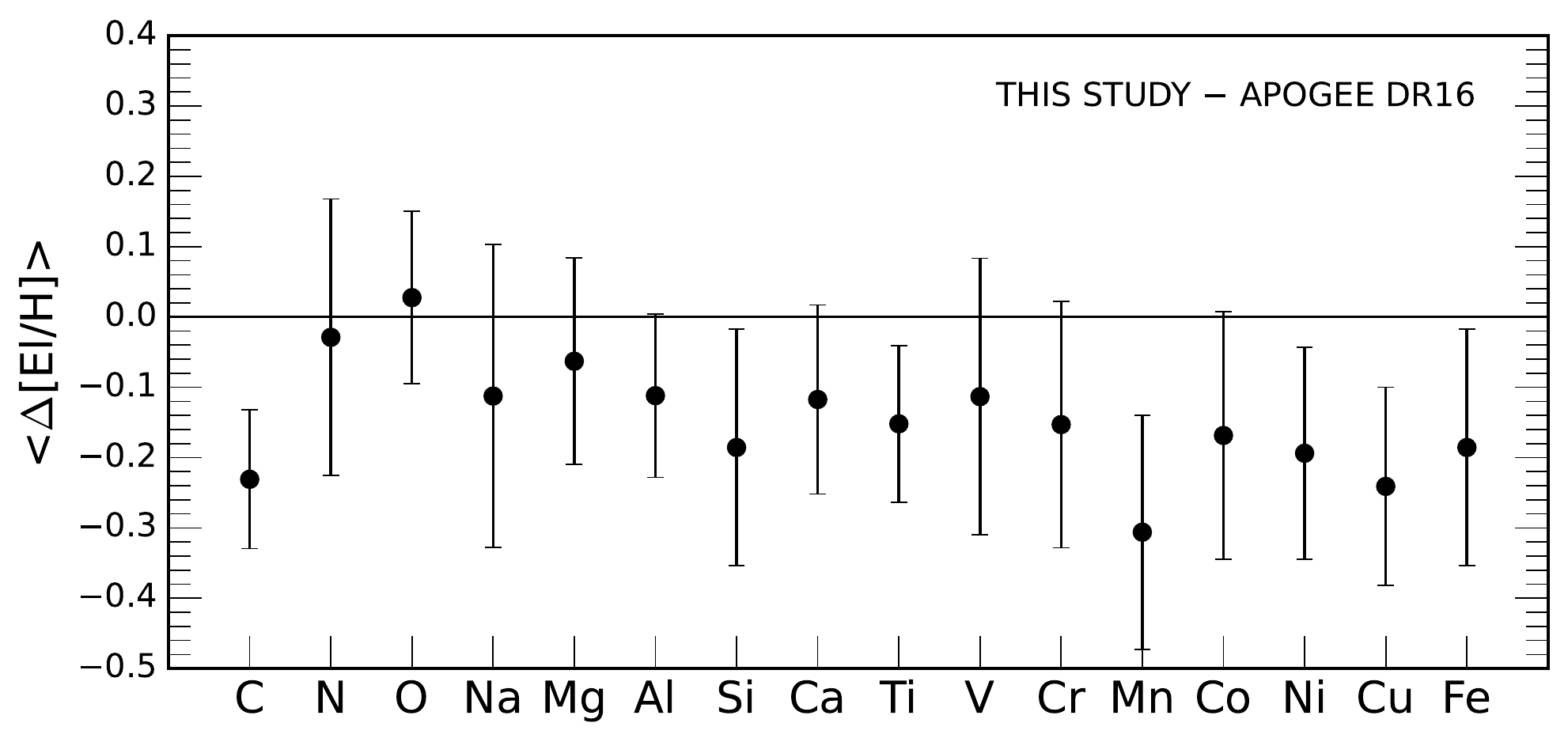}
\caption{Comparison of abundances for the 55 stars in common with APOGEE DR16.
The average differences and standard deviations are calculated as our results minus the comparison.}
\label{fig:Oour_Apogee}
\end{figure}

Our derived main atmospheric parameters and chemical abundances of many elements complement other large spectroscopic catalogs of, e.g., \citet{DelgadoMena2021} and \citet{Bensby2014} among others, because we do not have stars in common with these high-resolution spectroscopic catalogs. We have only one star in common in our study and that of \citet{Bensby2014}, and the parameters of this star agree well.

In the compilation of the PASTEL catalog (\citealt{Soubiran2016}), we found few studies that have derived stellar parameters for some stars of our sample. PASTEL is a bibliographical catalog that compiles determinations of stellar atmospheric parameters ($T_{\rm eff}$, log\,$g$, [Fe/H]) and provides determinations obtained from detailed analyses of high resolution spectra with nigh S/N. It provides atmospheric parameters derived from various methods.
We have 107 stars in common that were observed before with derived main atmospheric parameters that are collected in the PASTEL catalog.
The consistency of the effective temperature and surface gravity in our sample and PASTEL (catalog version of 2020 January 30 as in VizieR) is quite good: $\langle \Delta T_{\rm eff} \rangle=36 \pm 92$~K and $\langle \Delta {\rm log}~g \rangle=0.15 \pm 0.41$. The scatter of $\Delta{\rm log}\,g$ is caused by the variety of surface gravity determination methods (e.g. different line lists) used for the values presented in the PASTEL compilation.

We found 55 stars in common with the 16th data release (DR16) of the near-infrared, large-scale, stellar spectroscopic survey APOGEE \citep{Jonsson20}. For the comparison of the two studies, we used the calibrated parameters and abundances determined with the APOGEE Stellar Parameters and Chemical Abundance Pipeline (ASCAP, version l33, release 12; \citealt{GarciaPerez16}). The biases for the main stellar atmospheric parameters from our sample are $\langle \Delta T_{\rm eff} \rangle=-23 \pm 93$~K and $\langle \Delta {\rm log}~g \rangle=0.07 \pm 0.29$~dex. Having in mind  a complex approach of calibration was adopted for the surface gravity determinations in the APOGEE survey (see \citealt{Jonsson20} Section 5.2.2 for more details), the agreement between our results is rather good.

In Figure~\ref{fig:Oour_Apogee} we show our comparison of [Element/H] abundances for the 55 stars in common with APOGEE DR16.
The average differences for all stars in common and the standard deviations are calculated as our results minus the comparison. The sample of common stars is not very large; however, as one can see, the majority of elements have a systematic shift of about $-0.15$~dex.  
We refer the reader to \citet{Jonsson20} and their Table~12 where APOGEE~DR16 results were compared with independent high-resolution optical spectroscopic works. It was found that DR16 also has an average shift of about $-0.05$~dex compared to the other optical studies.

Regarding the C, N, and O elements, the APOGEE survey DR16 uses the infrared lines of CO, CN, and OH molecules, respectively, and the agreement for the nitrogen and oxygen abundances is quite good. The abundances of carbon, on the other hand, differ by about 0.2~dex. The larger  carbon abundances determined in the APOGEE survey could be caused by the weakness of the investigated CO bands and by difficulties in the placement of the continuum.

\section{Star--Planet Connections}

\begin{table*}
\centering
\renewcommand{\tabcolsep}{0.95mm}
 \caption{Stars of Our Sample with Confirmed Exoplanets.}
 \label{tab:exoplanets}
 \begin{tabular}{lcccccllc}
  \hline
  \hline
  { \raisebox{-1.5ex}[0cm][0cm]{TYC ID}} & { \raisebox{-1.5ex}[0cm][0cm]{Planet}} & Planet Mass & Planet Mass & Orbital Period & Ref.$^{(1)}$ &  \raisebox{-1.5ex}[0cm][0cm]{C/O} & \raisebox{-1.5ex}[0cm][0cm]{Mg/Si}& \raisebox{-1.5ex}[0cm][0cm]{[\ion{Fe}{1}/H]}\\

 & &($M_\mathrm{Earth}$)& ($M_\mathrm{Jupiter}$) & (days) &  &  & \\
  \hline
2099-2717-1	&	HD 164922 b	&	$116_{-12}^{\text{+}10}$	&	$0.365_{-0.038}^{\text{+}0.031}$	&		$1207_{-5}^{\text{+}4}$		&	1	&	0.59$^a$	&	1.44$^b$	&	0.17$^b$	\\
	&	HD 164922 c	&	$13_{-2}^{\text{+}3}$	&	$0.041_{-0.006}^{\text{+}0.009}$	&		$75.74_{-0.05}^{\text{+}0.06}$		&	1	&		&		&		\\
	&	HD 164922 d	&	$4\pm{1}$	&	$0.01\pm{0.00}$	&		$12.458\pm{0.003}$		&	1	&		&		&		\\
	&	HD 164922 e	&	$10.5\pm{1.0}$	&	$0.0331\pm{0.0031}$	&		$41.763\pm{0.012}$		&	2	&		&		&		\\
 \noalign{\smallskip}																			
2103-1620-1	&	HD 164595 b	&	$16.14\pm{2.72}$	&	$0.05078\pm{0.00856}$	&		$40\pm{0.24}$		&	3	&	...	&	1.05$^b$	&	$-$0.13$^b$	\\
2595-1464-1	&	HD 155358 b	&	$315\pm{25}$	&	$0.99\pm{0.080}$	&		$194.3\pm{0.30000}$		&	4	&	0.26$^a$	&	1.55$^b$	&	$-$0.66$^b$	\\
	&	HD 155358 c	&	$261\pm{22}$	&	$0.82\pm{0.07}$	&		$391.9\pm{1.0}$		&	5	&		&		&		\\
 \noalign{\smallskip}																			
2648-2151-1	&	HD 178911 B b	&	$2552\pm{798}$	&	$8.03\pm{2.510}$	&		$71.484\pm{0.02000}$		&	4	&	...	&	1.08$^b$	&	0.20$^b$	\\
3067-576-1	&	14 Her b	&	$1481\pm{48}$	&	$4.66\pm{0.15}$	&		$1773.40002\pm{2.50000}$		&	4	&	...	&	1.60$^b$	&	0.33$^b$	\\
	&	HD 145675 c	&	$1843_{-318}^{\text{+}445}$	&	$5.8_{-1.0}^{\text{+}1.4}$	&		$25000_{-9200}^{\text{+}24000}$		&	2	&		&		&		\\
 \noalign{\smallskip}																			
3501-1373-1	&	HD 154345 b	&	$261\pm{22}$	&	$0.82\pm{0.07}$	&		$3341.55884\pm{93.00000}$		&	4	&	...	&	0.95$^b$	&	$-$0.13$^b$	\\
3525-186-1	&	HD 168009 b	&	$9.53_{-1.18}^{\text{+}1.21}$	&	$0.03_{-0.0037}^{\text{+}0.0038}$	&		$15.1479_{-0.0037}^{\text{+}0.0035}$		&	2	&	0.62$^a$	&	1.23$^b$	&	$-$0.01$^b$	\\
3565-1525-1	&	16 Cyg B b	&	$566\pm{25}$	&	$1.78\pm{0.08}$	&		$798.5\pm{1.00000}$		&	4	&	0.55	&	1.15	&	$-$0.05	\\
3869-494-1	&	HD 139357 b	&	$3101.89\pm{683.306}$	&	$9.76\pm{2.150}$	&		$1125.7\pm{9.000}$		&	6	&	0.3	&	1.09	&	0.23	\\
3875-1620-1	&	iot Dra b	&	$2803.1\pm{228.8}$	&	$8.82\pm{0.72}$	&		$511.098\pm{0.089}$		&	7	&	0.32	&	0.99	&	$-$0.05	\\
3888-1886-1	&	HD 158259 b	&	$2.22_{-0.45}^{\text{+}0.39}$	&	$0.00698_{-0.00142}^{\text{+}0.00123}$	&		$2.178_{-0.00010}^{\text{+}0.00009}$		&	8	&	0.61$^a$	&	1.17$^b$	&	$-$0.07$^b$	\\
	&	HD 158259 c	&	$5.6_{-0.59}^{\text{+}0.60}$	&	$0.0176_{-0.0019}^{\text{+}0.0019}$	&		$3.432_{-0.00016}^{\text{+}0.00030}$		&	8	&		&		&		\\
	&	HD 158259 d	&	$5.41_{-0.71}^{\text{+}0.74}$	&	$0.017_{-0.0022}^{\text{+}0.0023}$	&		$5.1980814_{-0.0008814}^{\text{+}0.0008186}$		&	8	&		&		&		\\
	&	HD 158259 e	&	$6.08_{-1.03}^{\text{+}0.94}$	&	$0.0191_{-0.0032}^{\text{+}0.0030}$	&		$7.951_{-0.0021}^{\text{+}0.0022}$		&	8	&		&		&		\\
	&	HD 158259 f	&	$6.14_{-1.37}^{\text{+}1.31}$	&	$0.0193_{-0.0043}^{\text{+}0.0041}$	&		$12.028\pm{0.009}$		&	8	&		&		&		\\
 \noalign{\smallskip}																			
3903-2143-1	&	HD 167042 b	&	$540.29_{-38.14}^{\text{+}28.60}$	&	$1.7_{-0.12}^{\text{+}0.9}$	&		$420.77_{-3.11}^{\text{+}3.48}$		&	9	&	0.32	&	1.05	&	$-$0.01	\\
3910-257-1	&	HD 163607 b	&	$249.1\pm{3.1}$	&	$0.7836\pm{0.0098}$	&		$75.2203\pm{0.0094}$		&	10	&	0.50$^c$	&	1.17$^c$	&	0.17$^c$	\\
	&	HD 163607 c	&	$699.5\pm{11.8}$	&	$2.201\pm{0.037}$	&		$1272\pm{4.4}$		&	10	&		&		&		\\
 \noalign{\smallskip}																			
4222-2311-1	&	42 Dra b	&	$1233.13\pm{270.14}$	&	$3.88\pm{0.85}$	&		$479.1\pm{6.2}$		&	6	&	0.28$^c$	&	1.23$^c$	&	$-$0.47$^c$	\\
4412-1654-1	&	HD 143105 b	&	$385\pm{19}$	&	$1.21\pm{0.06}$	&		$2.1974\pm{0.0003}$		&	11	&	0.51	&	1.15	&	$-$0.07	\\
4414-2315-1	&	11 UMi b	&	$4685\pm{795}$	&	$14.74\pm{2.50}$	&		$516.21997\pm{3.20000}$		&	4	&	0.24	&	0.95	&	$-$0.34	\\
4416-1799-1	&	bet UMi b	&	$1938.7\pm{317.8}$	&	$6.1\pm{1.0}$	&		$522.3\pm{2.7}$		&	12	&	0.19	&	1.03	&	$-$0.50	\\
4417-267-1	&	8 UMi b	&	$416\pm{51}$	&	$1.31\pm{0.16}$	&		$93.4\pm{4.50000}$		&	4	&	0.14	&	1.32	&	$-$0.16	\\
4436-1424-1	&	psi 1 Dra B b	&	$486\pm{32}$	&	$1.53\pm{0.10}$	&		$3117\pm{42}$		&	13	&	0.55$^a$	&	1.29$^b$	&	$-$0.08$^b$	\\
4494-1346-1	&	HD 7924 b	&	$6.4\pm{0.00}$	&	$0.02\pm{0.00}$	&		$5.39792\pm{0.00025}$		&	4	&	0.44$^a$	&	1.23$^b$	&	$-$0.30$^b$	\\
	&	HD 7924 c	&	$7.86_{-0.71}^{\text{+}0.73}$	&	$0.0247_{-0.0022}^{\text{+}0.0023}$	&		$15.299_{-0.0033}^{\text{+}0.0032}$		&	14	&		&		&		\\
	&	HD 7924 d	&	$6.44_{-0.78}^{\text{+}0.79}$	&	$0.0203_{-0.0025}^{\text{+}0.0025}$	&		$24.451_{-0.017}^{\text{+}0.015}$		&	14	&		&		&		\\
 \noalign{\smallskip}																			
4532-2096-1	&	HD 33564 b	&	$2892.1$	&	$9.1$	&		$388\pm{3}$		&	15	&	...	&	1.11$^b$	&	$-$0.16$^b$	\\
4561-2319-1	&	HD 120084 b	&	$1430.2$	&	$4.5$	&		$2082_{-35}^{\text{+}24}$		&	16	&	0.30	&	1.28	&	$-$0.07	\\
4575-1336-1	&	HD 150706 b	&	$861.28_{-209.76}^{\text{+}362.31}$	&	$2.71_{-0.66}^{\text{+}1.14}$	&		$5894_{-1498}^{\text{+}5584}$		&	17	&	0.45$^a$	&	1.12$^b$	&	$-$0.15$^b$	\\
4576-1486-1	&	HD 158996 b	&	$4450\pm{731}$	&	$14\pm{2.3}$	&		$820.2\pm{14.0}$		&	18	&	0.19	&	0.97	&	$-$0.47	\\
4650-917-1	&	HD 216520 b	&	$10.26\pm{0.99}$	&	$0.03228\pm{0.00311}$	&		$35.45\pm{0.011}$		&	19	&	0.40$^a$	&	1.10$^b$	&	$-$0.35$^b$	\\
	&	HD 216520 c	&	$9.44\pm{1.630}$	&	$0.0297\pm{0.00513}$	&		$154.43\pm{0.44}$		&	19	&		&		&		\\
 
  \hline
 \end{tabular}
 \flushleft
        {\bf Note.} Some host C, O, Mg and Si elemental abundances are taken from our previous works: 
       $^a$ \citet{Stonkute2020},
       $^b$ \citet{Mikolaitis2019},
       $^c$ \citet{Tautvaisiene20}.\\
       {\bf{References.}} $^{(1)}$ Planet mass and period references: 1 - \citet{Benatti20}, 2 - \citet{Rosenthal21}, 3 - \citet{Courcol15}, 4 - \citet{Stassun17}, 5 - \citet{Robertson12}, 6 - \citet{Dollinger09}, 7 - \citet{Butler06}, 8 - \citet{Hara20}, 9 - \citet{Bowler10}, 10 - \citet{Luhn19}, 11 - \citet{Hebrard16}, 12 - \citet{Lee14}, 13 - \citet{Endl16}, 14 - \citet{Fulton15}, 15 - \citet{Galland05}, 16 - \citet{Sato13}, 17 - \citet{Boisse12}, 18 - \citet{Bang18}, 19 - \citet{Burt21}. 
 \label{table:exoplanets}
\end{table*}

A list of 25 planet-hosting stars with their C/O, Mg/Si abundance ratios, [\ion{Fe}{1}/H] and information about their planets is presented in Table~\ref{table:exoplanets}. The planetary mass ($M_{p}$sin${i}$) and orbital period (in days) were taken from the NASA Exoplanet Archive on 2021 December 15. The references are provided in Table~\ref{table:exoplanets}. Stars HD~158259, HD~7924, and HD~21652 host five, three, and two low-mass planets, respectively. Star HD~164922 has three low-mass and one high-mass planets. Three more stars have two confirmed high-mass planetary systems. According to \citet{Kokaia2020}, the stars HD~154345 and HD~150709 may be candidate systems with high resilient habitability and have good prospects for the detection of planets in their habitable zones. According to \citet{Agnew2018}, $\psi$\,1\,Dra also has as high as 0--25\% probability of resilient habitability. In our sample, the two planet-hosting stars HD~155358 and HD~145675 belong to the thick disk of the Galaxy (one belongs to the metal-poor and one to the metal-rich parts), and all the remaining stars are the thin-disk stars. 

\subsection{Stellar Chemical Composition and Planet Mass Relation}

\begin{figure}

  \includegraphics[width=\columnwidth]{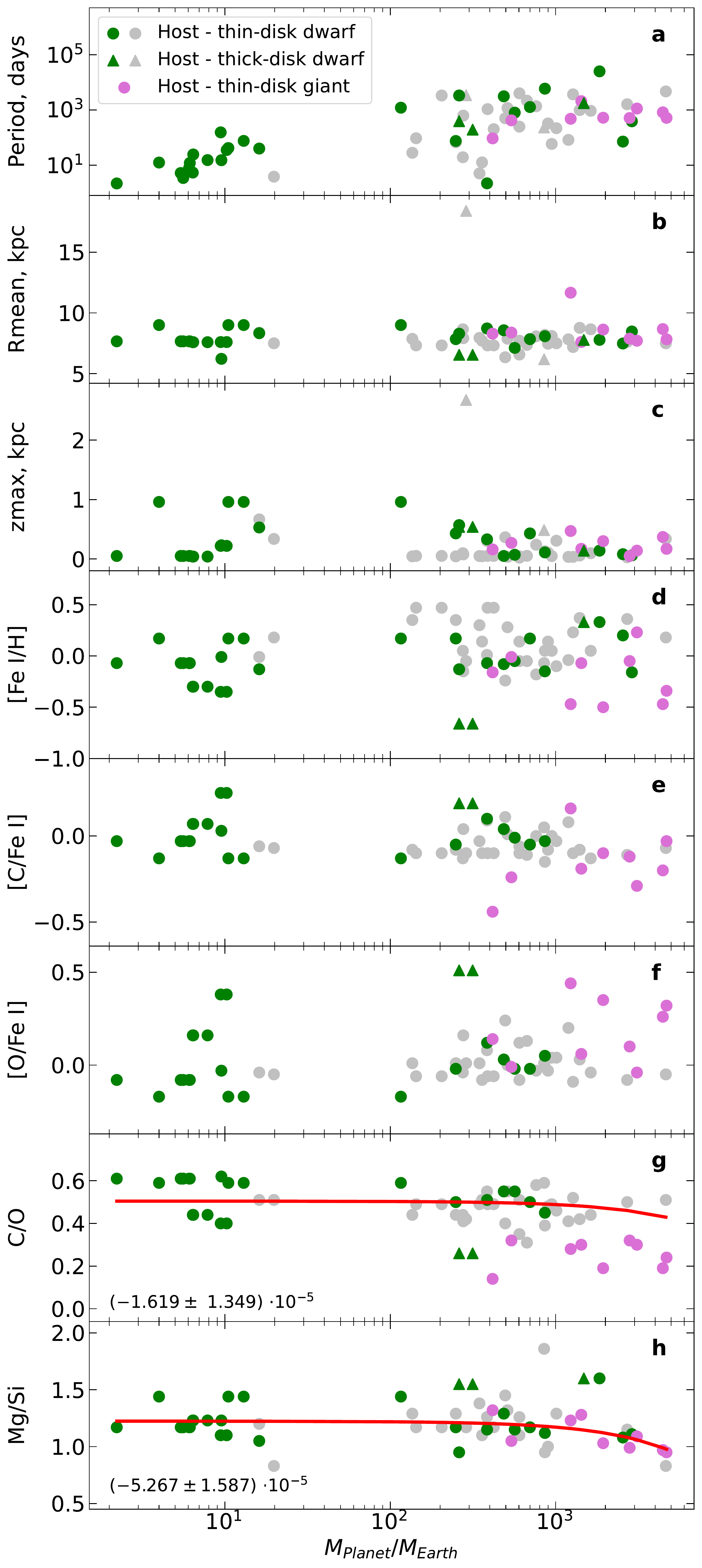}
\caption{Distributions of parameters and elemental abundances in planet-hosting stars and planets as a function of planet masses. 
The green symbols indicate the planet-hosting dwarfs, the pink symbols are for giants, and the gray symbols are for dwarfs investigated by \citet{Mishenina2021}. The circles indicate thin-disk stars, and the triangles are for thick-disk stars. See the text for more information.}
\label{fig:param-planet-mass}
\end{figure}

Figure~\ref{fig:param-planet-mass} displays distributions of parameters and elemental abundances  in  planet-hosting  stars  and  planets  as  a  function of  planet  masses.  We marked planets orbiting dwarfs and giants as well as thin- and thick-disk stars with different symbols. Along with our sample, we also display thr results from the recent study by \citet{Mishenina2021}. We updated the parameters of planets in their study according to the NASA Exoplanet Archive data of 2021 December 15 and computed $R_{\rm mean}$ and $z_{\rm max}$ for their host stars. All the investigated stars in that study are thin-disk dwarfs. 

According to the mass, the exoplanets fall into two widely separated mass ranges (Figure~\ref{fig:param-planet-mass}a). One group of planets has masses from 2.22 to 16.14~$M_{\rm Earth}$ (we call them low-mass planets), and another group is from 116 to 3102~$M_{\rm Earth}$ (high-mass planets).  

The values of $R_{\rm mean}$ in our sample of planet-hosting stars range from 6.22 to 9~kpc, except for one star that is located at 11.66~kpc. The $z_{\rm max}$ values are up to 0.57~kpc, except for one star with four planets, which is at 0.96~kpc. 

When we compare the metallicity (Figure~\ref{fig:param-planet-mass}d), it is on average higher by about 0.2~dex in dwarfs with confirmed high-mass planets than in dwarfs with low-mass planets. This is in agreement with findings by \citet{Adibekyan2012}. 
However, there are giant stars of lower metallicity with high-mass planets, but we do not have giants with confirmed low-mass planets in our sample for a comparison. A comprehensive review of the metallicity of planet-hosting stars can be found in \citet{Adibekyan2019}.

\begin{figure}
  \centering
  \includegraphics[width=\columnwidth]{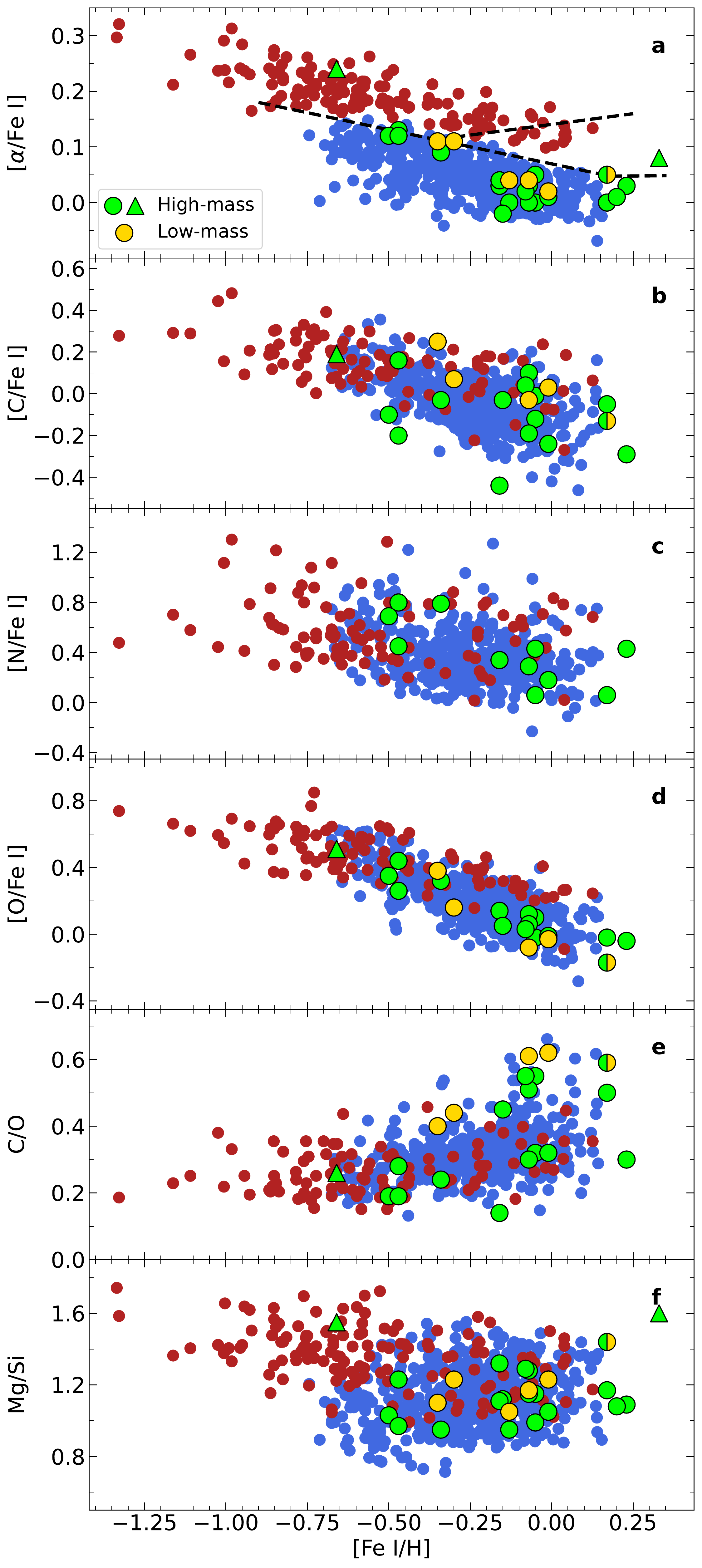}
  \caption{Elemental abundance ratios as a function of metallicity. The blue dots represent the thin-disk stars, the red dots show the thick-disk. The dashed lines that separate the disks are from \citet{Lagarde2021}. The high-mass planet-hosting stars are marked by green symbols (the two belonging to the thick disk are marked by triangles), the stars with low-mass planets are displayed by yellow symbols, and the one with both one high-mass planet and three low-mass planets is shown by the circle filled with both colors.}
  \label{fig:Planet-metal}
\end{figure}

We also searched for signatures of different carbon and oxygen abundances in low- and high-mass planet-hosting stars. We found no significant correlation between carbon or oxygen abundances and planetary masses. The same conclusion concerning carbon was reached by \citet{Suarez2017}. 

Of course in giants, the carbon abundances are lower by about 0.2~dex than in dwarfs due to material mixing effects in evolved stars (see \citealt{Tautvaisiene2010}), and the oxygen abundances are larger due to the lower stellar metallicity and the corresponding Galactic chemical evolution results, and the C/O ratios, consequently, are lower by about 0.25. 

We computed the linear fits for the C/O and Mg/Si versus planetary mass similar to \citet{Mishenina2021}, and we  used their data to compliment ours (panels (g) and (h) in Figure~\ref{fig:param-planet-mass}). Our sample has more dwarfs with low-mass planets, while with complementary giants, we added more stars with high-mass planets to the sample, which can be used to compute the Mg/Si slope. We find a weak negative C/O slope with a Pearson correlation coefficient (PCC) equal to $-0.17$ and a slightly more negative Mg/Si slope with PCC$=-0.37$ toward the stars with high-mass planets. The Mg/Si versus planetary mass slope is exactly the same as it was found by \citet{Mishenina2021}. \citet{Adibekyan2015} also suggested that low-mass planets are more prevalent around stars with high Mg/Si ratio. \citet{Suarez18} also found a slightly negative slope for Mg/Si, but inferred a positive slope for C/O versus planetary mass.

We also looked for similarities or differences between the stars without confirmed planets and the low- and high-mass planet-hosting stars.
The abundance ratios of Mg/Si, C/O and the element abundances [C/Fe], [N/Fe], and [O/Fe] together with $\alpha$-elements as a function of metallicity are presented in Figure~\ref{fig:Planet-metal}.  The thin- and thick-disk stars are displayed with different colors, as are the low- and high-mass planet-hosting stars. The [$\alpha$/Fe] values are computed from the abundances of Mg, Si, Ca, and Ti elements. 

\citet{Adibekyan2012b} found that the [El/Fe]
ratios for $\alpha$-elements both for high- and low-mass planet hosts are systematically higher than those of comparison stars at low metallicities ($\rm{[Fe/H]}\le -0.2$). The stars in our sample confirm this finding (Figure~\ref{fig:Planet-metal}a). 

We found that both planet hosts and non-planet hosts have similar Mg/Si ratios and CNO abundances. We agree with the study by \citet{Bedell2018} that the ratios of carbon-to-oxygen and magnesium-to-silicon in solar-metallicity stars are homogeneous,  implying that exoplanets may exhibit rather small diversity of composition. However,  in our sample we can see (the yellow symbols in Figure~\ref{fig:Planet-metal}e) that slightly higher C/O ratios seem to be present in the dwarfs hosting low-mass planets. 
\citet{DelgadoMena2021} found tentative evidence that stars with low-mass planets have higher [C/Fe] ratios at lower metallicities than stars without planets at the same metallicity.

The cosmic distribution of Mg/Si for the sample stars is broader than that of C/O. \citet{DelgadoMena2010}  found that Mg/Si abundance ratios are higher in stars without confirmed planets. This is not evident in our sample (Figure~\ref{fig:Planet-metal}f). 

The work on elemental abundances in planet-hosting stars has to be continued. Investigations of CNO abundances are especially challenging as there are few spectral features for a robust abundance determination. An encouraging attempt to use the NH band at 3360~\AA\ was performed by \citet{Suarez-Andres2016}. In this study, the nitrogen distributions for stars with and without planets show that planet hosts are nitrogen rich when compared to single stars. However, it was remarked that given the linear trend between [N/Fe] versus [Fe/H], this fact can be explained as being due to the metal-rich nature of planet hosts. In our study, the nitrogen abundances were determined for 11 high-mass planets hosting-stars, 9 of which are giants and 2 are dwarfs, and no concrete conclusions can be drawn so far.   

\subsection{Elemental Abundances versus Condensation Temperatures}

\begin{figure*}
 \centering
\includegraphics[width=0.99\textwidth]{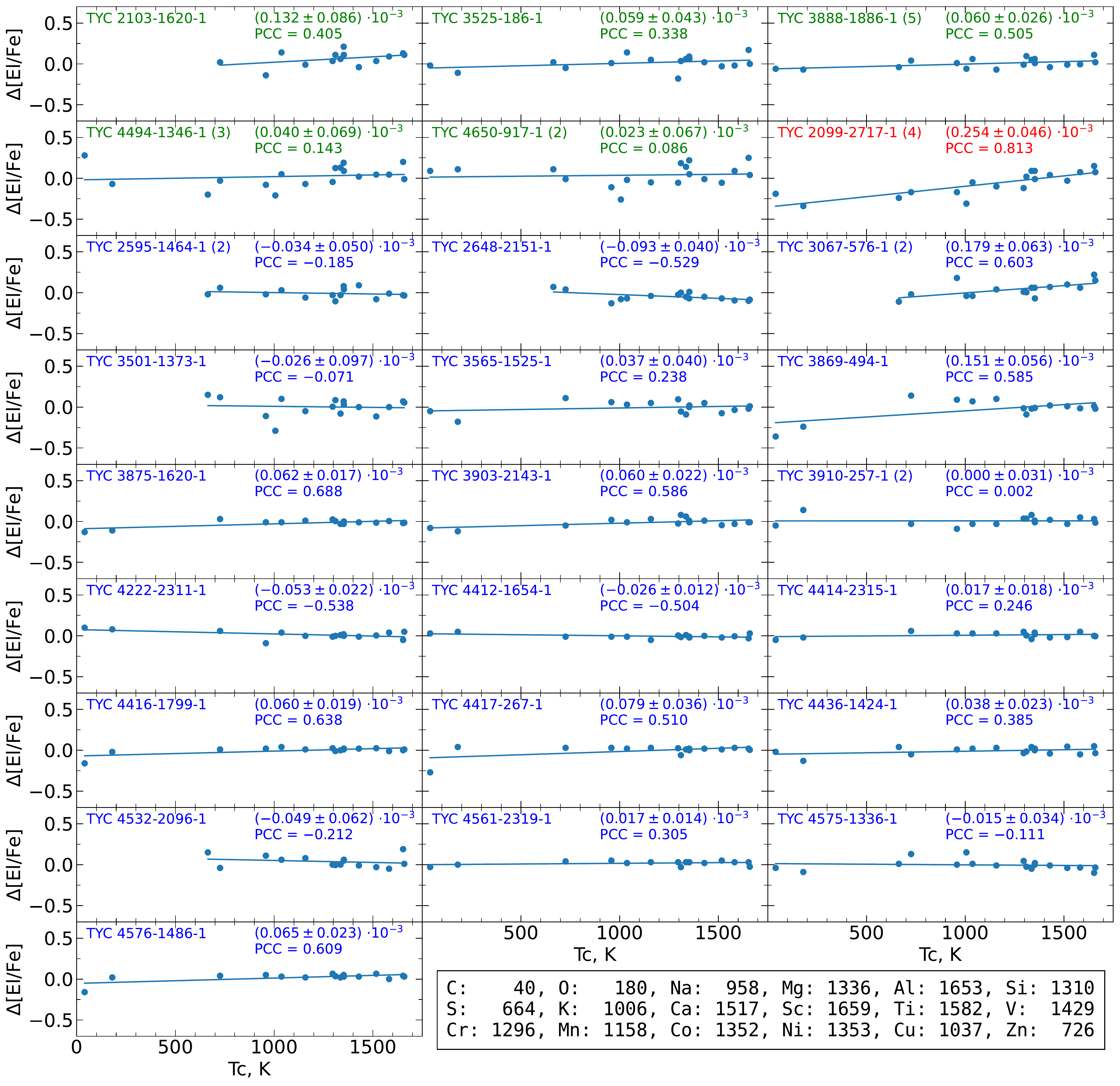}
\caption{Differences between elemental abundances in the planet-hosting and comparison stars $\Delta$[El/Fe] as a function of the condensation temperatures of chemical elements. The green labels mark stars hosting low-mass planets, the blue labels show the stars with high-mass planets, and the red label marks the star with both types of planets. For stars hosting more than one confirmed planet, their number is provided in brackets near the name. The Pearson correlation coefficients and the slopes for the linear regression analysis are displayed in the upper-right corners of the plots.  }
\label{fig:kondensacijos}
\end{figure*}

Differences between elemental abundances in the planet-hosting and comparison stars $\Delta$[El/Fe] as a function of the condensation temperatures of chemical elements is another open topic of discussion. Many studies have reported volatile and refractory element abundance variations in planet-hosting stars and their dependence on the elemental condensation temperatures. This question was addressed for the Sun as one of  the first planet-hosting stars because the element abundances can be determined with  very high accuracy. \citet{Melendez2009} found that the Sun shows a characteristic signature with a $\sim20$\% depletion of refractory elements relative to the volatile elements in comparison with the solar twins and that the abundance differences correlate strongly with the condensation temperatures of the elements. However, comparisons of stars in binary systems in which only one of the stars has a detected planet show either an opposite pattern (e.g. \citealt{TucciMaia2014, Saffe2019}) or negligible slopes (e.g. in  \citealt{Liu2021}). 

\citet{GonzalezHernandez2013} in their study of 29 planet-hosting stars found that solar-type stars with only giant planets with long orbital periods display mostly negative slopes. In the work by \citet{Liu2020}, a sample of 16 planet-hosting solar-type stars exhibited a variety of abundance--$T_{\rm c}$ trends with no clear dependence upon age, metallicity, or $T_{\rm eff}$. In a sample of 25 planet-hosting dwarf stars, \citet{Mishenina2021} inferred a possible relation between the negative slope and planetary masses. Studies of this topic clearly need more precisely determined data.   

Because the investigation of elemental abundance correlation with $T_{\rm c}$ needs comparison objects, it is difficult to select a proper comparison object of the same age, location in the Galaxy, and atmospheric parameters for single stars that could allow accounting for the stellar and Galactic chemical evolution. 
Together with our previous studies (\citetalias{Tautvaisiene20}, \citealt{Mikolaitis2018, Mikolaitis2019, Stonkute2020}), the number of our homogeneously analyzed stars has risen to 1266. This gives us the possibility of selecting proper comparison stars to answer the question whether planet-hosting stars show an abundance that is different from that of stars without identified planets. As all the stars were studied homogeneously, we are mostly save from any biases or offsets that could occur when comparing stars. 
We identified one counterpart for only one planet-hosting star, while for others, up to 13 stars with similar parameters were found. In this case, the mean of all similar stars was taken for comparison. When we searched for similar objects, we aimed for a difference in the $T_{\rm eff}$, log\,$g$, [Fe/H], and $v_t$ values of no more than their determination uncertainties, as well as close $R_{\rm mean}$, $z_{\rm max}$, and age values. We took into account the dependence on to the Galactic disks as well. 

In Figure~\ref{fig:kondensacijos} we show the differences between elemental abundances $\Delta$[El/Fe] in the planet-hosting and comparison stars as a function of the condensation temperatures of chemical elements. We mark stars hosting low- and high-mass planets with colored labels, and also mark the one star that hosts both types of planets (red label). It is easy to notice that the last one shows the largest positive slope.  For stars hosting more than one confirmed planet, their number is provided in brackets near the name. 
The Pearson correlation coefficients and the slopes for the linear regression analysis are displayed in the upper-right corners of the plots. 

\begin{figure}
  \centering
  \includegraphics[width=\columnwidth]{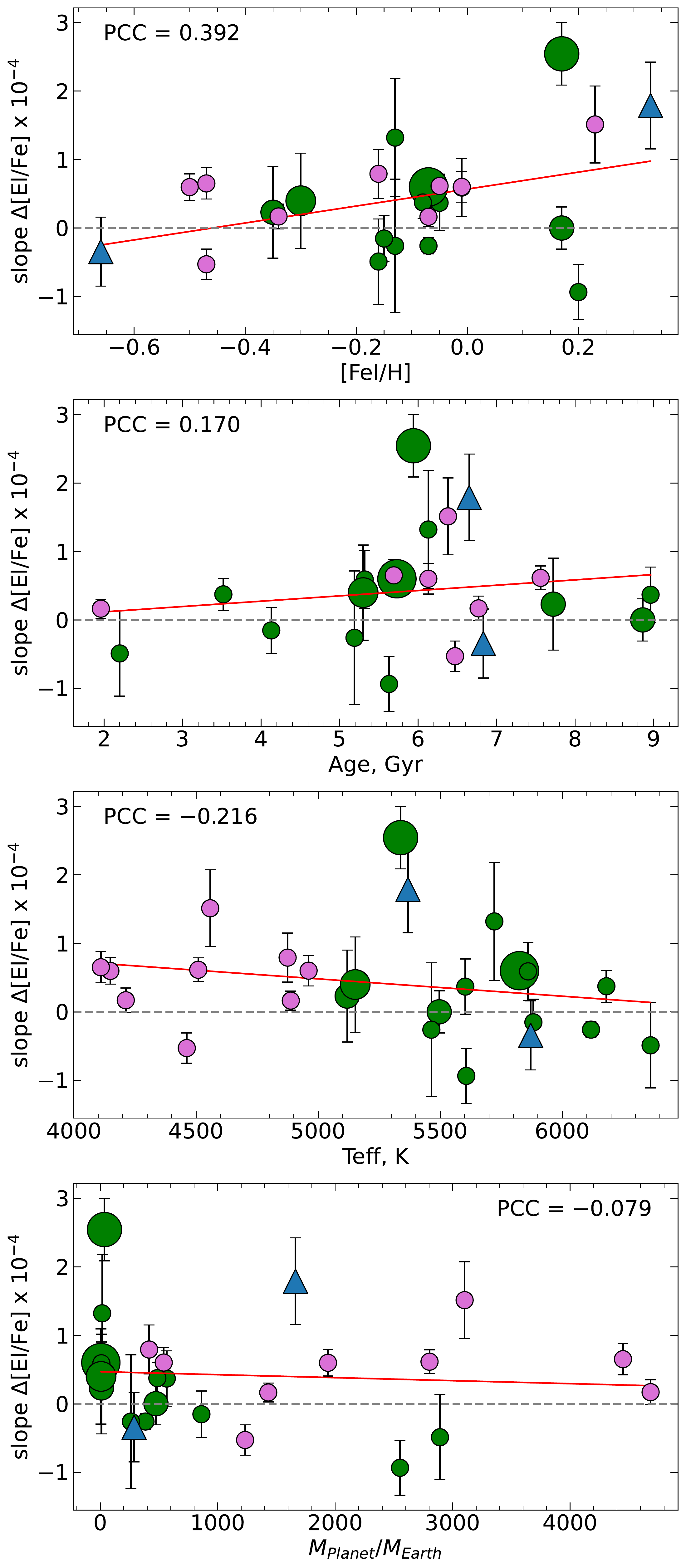}
  \caption{Dependences of $\Delta{\rm [El/Fe]}-T_{\rm c}$ slopes on the stellar parameters and planet masses. The green circles represent the thin-disk dwarfs, the pink dots show the thin-disk giants, and the thick-disk dwarfs are marked by blue triangles. The size of symbols depend on the number of confirmed planets in the system. The red lines display linear fits to the data points, and their Pearson correlation coefficients are presented as labels in the corners of the panels.}
  \label{fig:kondens-param}
\end{figure}

In Figure~\ref{fig:kondens-param} we show dependences of $\Delta{\rm [El/Fe]}-T_{\rm c}$ slopes on the stellar metallicity, age, effective temperature, and planet masses. For stars with several planets, we took their averaged masses. The  thin-disk dwarfs and giants as well as the thick-disk dwarfs are marked with different symbols. The size of symbols represents the number of planets. The red lines display linear fits to the data points, and their Pearson correlation coefficients are presented as labels in the corners of the panels.

All the low-mass planet hosts in our sample show positive slopes in $\Delta$[El/Fe] versus $T_{\rm c}$ planes, in particular, the star with the large number of various planets.  Looking at the high-mass planet-hosts we see the diversity of slopes, but our results mean that in more metal rich, older, and cooler stars, the positive elemental abundance slopes are more common.

We refer to comprehensive theoretical studies concerning possible scenarios of planet formation and star--planet chemical composition relations, which try to incorporate and explain the observational results (e.g. \citealt{ Thiabaud15, Bitsch20, Adibekyan2021, Cowley2021, Nibauer2021, Schneider21, Schulze2021}, and references therein). 

\section{Summary}

In this paper, we extended the determination of the main atmospheric parameters, ages, kinematic parameters, and abundances of up to 24 chemical species from  high-resolution spectroscopy of bright, ($V<8$~mag), slowly rotating stars cooler than F5 spectral type in a circle up to about $12^{\circ}$ surrounding the northern TESS CVZ.

A detailed characterization was done for a sample of \noatmospheric ~stars of different evolutionary stages, ages, and atmospheric parameters:
$T_{\rm eff}$ varied between 3980 and 6400~K, and [Fe/H] varied between $-2.25$ and 0.15~dex. The stellar ages varied from 0.2 to 12~Gyr. A distinctive ${\rm log}~g$ distribution clearly separated giant and dwarf stars; the parameter of the former  varied between 0.5 and 3.5, with a mean at 2.4, while the latter displayed values between 3.6 and 4.5 with a mean at 4.0.
 Data from the Gaia EDR3 catalog were used to calculate the stellar kinematic parameters. The mean galactocentric distances, $R\mathrm{_{mean}}$, span from 4 to 10~kpc, and the distances from the Galactic plane, $z\mathrm{_{max}}$, reach 2.4~kpc with the $<z\mathrm{_{max}}>=0.30 \pm{0.26}$~kpc. Stellar velocity components  (\textit{U, V}, and \textit{W}) were determined as well.

Using a sample of 25 planet-hosting stars, we investigated the stellar chemical composition and planet mass relation, compared elemental abundances with stars without detected planets, and verified elemental abundance versus condensation temperature slopes. 

The sample contains stars with five, three, and two low-mass planets. One star has three low-mass planets and one high-mass planet, and three more stars have two confirmed high-mass planetary systems. Three stars may be candidate systems with high resilient habitability and have good prospects for the detection of planets in their habitable zones. Two planet-hosting stars belong to the thick disk of the Galaxy (one belongs to the metal-poor and one to the metal-rich parts), and all the remaining stars are thin-disk stars. 

One group of stars hosts planets with masses from 2.22 to 16.14~$M_{\rm Earth}$ (we call them low-mass planets), and another group is from 116 to 3102~$M_{\rm Earth}$ (high-mass planets). The values of $R_{\rm mean}$ in our sample of planet-hosting stars range from 6.22 to 9~kpc, except for one star that is located at 11.66~kpc. The $z_{\rm max}$ values are up to 0.57~kpc, except for one star with four planets that is at 0.96~kpc. 

The analysis of planet-hosting stars in our sample drove us to the following conclusions:

\begin{itemize}
    \item 
The dwarf stars hosting high-mass planets are more metal rich than those with low-mass planets.
\item
We find a weak negative C/O slope with PCC=$-0.17$ and a slightly more negative Mg/Si slope with PCC$=-0.37$ toward the stars with high-mass planets.  
\item
The element-to-iron ratios for $\alpha$-elements for high- and low-mass planet hosts are systematically higher than those of comparison stars at lower metallicities.
\item
We found that both planet hosts and non-planet hosts have  similar  Mg/Si  ratios  and  CNO  abundances,  but slightly  higher  C/O  ratios seem  to  be  present  in  dwarfs  hosting  low-mass planets.  
\item
All the low-mass planet hosts in our sample show positive slopes in $\Delta$[El/Fe] vsersus $T_{\rm c}$ planes, in particular, the star with the large number of various planets. The high-mass planet hosts have a diversity of slopes.  
\item
Our results mean that in more metal rich, older, and cooler, stars the positive elemental abundance slopes are more common.  
\end{itemize}

In the rapidly expanding field of exoplanet research, the chemical composition results and information about stellar ages and birth locations determined in this work for stars in the northern hemisphere will provide useful priors in further exoplanet modelings as well as in Galactic evolution studies (see \citealt{Madhusudhan19, Madhusudhan2021a,  Unterborn19, Hinkel20, Clark21, Turrini2021}).

\vspace{5mm}

We acknowledge the grant from the European Social Fund via the Lithuanian Science Council (LMTLT) grant No. 09.3.3-LMT-K-712-01-0103. 
The authors would like to thank Dr.~Nikku~Madhusudhan for the helpful suggestions regarding this project. 
The anonymous referee is thanked for constructive comments. 
This research has made use of NASA’s Astrophysics Data System Bibliographic Services and SIMBAD database, operated at CDS, Strasbourg, France.
This work has made use of data from the European Space Agency
(ESA) mission Gaia (https://www.cosmos.esa.int/gaia), processed by
the Gaia Data Processing and Analysis Consortium (DPAC, https://www.
cosmos.esa.int/web/gaia/dpac/consortium). Funding for the DPAC has
been provided by national institutions, in particular the institutions participating
in the Gaia Multilateral Agreement.
This research has made use of the NASA Exoplanet Archive, which is operated by the California Institute of Technology, under contract with the National Aeronautics and Space Administration under the Exoplanet Exploration Program.
We are grateful to the Moletai Astronomical Observatory of Vilnius University 
for providing observing time for this project.

\vspace{5mm}

\facility{NASA Exoplanet Archive.}

\software{Astropy \citep{2018AJ....156..123A}, DAOSPEC \citep{Stetson2008}, MOOG \citep{Sneden1973}, galpy \citep{Bovy15},TURBOSPECTRUM \citep{Alvarez1998}, UniDAM \citep{Mints2017}.}

\appendix
\restartappendixnumbering 

\section{Appendix information}

Table~\ref{tab:CDS} lists the contents of the Machine-readable table (atmospheric parameters, kinematic properties, ages, and individual abundances) together with associated errors, and other information for the investigated stars. This table is available in its entirety in machine-readable form.

\begin{longtable}{lllll} 
\caption{Contents of the Machine-readable Table}\\
\hline
\hline
Col & Label & Units & Explanations \\
\hline
1	&	ID                  	&	  ---   	&	 Tycho catalog identification\\
2	&	TESS\_ID             	&	  ---   	&	 ID in the TESS catalog \\
3	&	Teff                	&	 K    	&	 Effective temperature\\
4	&	eTeff               	&	   K    	&	 Error on effective temperature\\
5	&	Logg                	&	  dex   	&	 Surface gravity\\
6	&	e\_Logg              	&	  dex   	&	 Error on surface gravity\\
7	&	[Fe/H]              	&	  dex   	&	 Metallicity \\
8	&	e\_[Fe/H]            	&	  dex   	&	 Error on metallicity \\
9	&	Vt                  	&	   km s$^{-1}$  	&	 Microturbulence velocity\\
10	&	e\_Vt                	&	  km s$^{-1}$   	&	 Error on microturbulence velocity\\
11	&	Vrad                	&	 km s$^{-1}$ 	&	 Radial velocity\\
12	&	e\_Vrad              	&	 km s$^{-1}$ 	&	 Error on radial velocity \\
13	&	Age                 	&	  log(yr)  	&	 Log age of the star \\
14	&	e\_Age               	&	  log(yr)  	&	 Error on log age\\
15	&	 U                  	&	  km s$^{-1}$  	&	 $U$ velocity\\
16	&	e\_U                 	&	  km s$^{-1}$   	&	 Error on $U$ velocity\\
17	&	V                   	&	 km s$^{-1}$    	&	 $V$ velocity \\
18	&	e\_V                 	&	 km s$^{-1}$    	&	 Error on $V$ velocity \\
19	&	W                   	&	 km s$^{-1}$    	&	 $W$ velocity \\
20	&	e\_W                 	&	 km s$^{-1}$     	&	 Error on $W$ velocity\\
21	&	 d                  	&	 kpc 	&	 Distance\\
22	&	R$_{\rm mean}$           	&	 kpc    	&	 Mean galactocentric distance\\
23	&	e\_R$_{\rm mean}$        	&	 kpc 	&	Error on mean galactrocentric distance\\
24	&	z$_{\rm max}$                	&	 kpc    	&	 Distance from Galactic plane\\
25	&	e\_z$_{\rm max}$              	&	 kpc    	&	 Error on distance from Galactic plane\\
26	&	{\it{e}}                   	&	 ---    	&	 Orbital eccentricity\\
27	&	e\_{\it{e}}                 	&	 ---  	&	 Error on orbital eccentricity\\
28	&	TD/D                	&	 --- 	&	 Thick- to thin-disk probability ratio\\
29	&	[C/H]    	&	 dex    	&	 Carbon abundance \\
30	&	e\_[C/H]  	&	 dex     	&	 Error on carbon abundance \\
... &  &  & \\
71	&	[\ion{Zn}{1}/H]    	&	 dex    	&	 Zinc abundance \\
72	&	e\_[\ion{Zn}{1}/H]  	&	 dex     	&	 Error on zinc abundance\\
73	&	[\ion{Fe}{1}/H]    	&	 dex    	&	  Iron abundance\\
74	&	e\_\ion{Fe}{1}/H]  	&	 dex    	&	 Error on iron abundance\\
75	&	[\ion{Fe}{2}/H]    	&	 dex    	&	 Ionized iron abundance \\
76	&	e\_[\ion{Fe}{2}/H]  	&	 dex     	&	 Error on ionized iron abundance\\
77  & [{alpha}/\ion{Fe}{1}] & dex &  Averaged \ion{Mg}{1}, \ion{Si}{1}, \ion{Si}{2}, \ion{Ca}{1}, \ion{Ca}{2}, \ion{Ti}{1}, and \ion{Ti}{2} to \ion{Fe}{1} abundance ratio \\
78  & e\_[{alpha}/\ion{Fe}{1}] & dex &  Standard error of the mean on [{alpha}/\ion{Fe}{1}] \\
79	&	Group  	&	 ---     	&	 Chemical attribution to the Galactic  subcomponent\\
\noalign{\smallskip}
\hline
\label{tab:CDS}
\end{longtable}

\bibliography{TESS}{}

\newcommand{\noop}[1]{}
\begin{thebibliography}{}
\expandafter\ifx\csname natexlab\endcsname\relax\def\natexlab#1{#1}\fi
\providecommand{\url}[1]{\href{#1}{#1}}
\providecommand{\dodoi}[1]{doi:~\href{http://doi.org/#1}{\nolinkurl{#1}}}
\providecommand{\doeprint}[1]{\href{http://ascl.net/#1}{\nolinkurl{http://ascl.net/#1}}}
\providecommand{\doarXiv}[1]{\href{https://arxiv.org/abs/#1}{\nolinkurl{https://arxiv.org/abs/#1}}}

\bibitem[{{Adibekyan}(2019)}]{Adibekyan2019}
{Adibekyan}, V. 2019, Geosciences, 9, 105, \dodoi{10.3390/geosciences9030105}

\bibitem[{{Adibekyan} {et~al.}(2017){Adibekyan}, {Gon{\c{c}}alves da Silva},
  {Sousa}, {Santos}, {Delgado Mena}, \& {Hakobyan}}]{Adibekyan2017}
{Adibekyan}, V., {Gon{\c{c}}alves da Silva}, H.~M., {Sousa}, S.~G., {et~al.}
  2017, Astrophysics, 60, 325, \dodoi{10.1007/s10511-017-9486-5}

\bibitem[{{Adibekyan} {et~al.}(2015){Adibekyan}, {Santos}, {Figueira}, {Dorn},
  {Sousa}, {Delgado-Mena}, {Israelian}, {Hakobyan}, \&
  {Mordasini}}]{Adibekyan2015}
{Adibekyan}, V., {Santos}, N.~C., {Figueira}, P., {et~al.} 2015, \aap, 581, L2,
  \dodoi{10.1051/0004-6361/201527059}

\bibitem[{{Adibekyan} {et~al.}(2021){Adibekyan}, {Dorn}, {Sousa}, {Santos},
  {Bitsch}, {Israelian}, {Mordasini}, {Barros}, {Delgado Mena}, {Demangeon},
  {Faria}, {Figueira}, {Hakobyan}, {Oshagh}, {Soares}, {Kunitomo}, {Takeda},
  {Jofr{\'e}}, {Petrucci}, \& {Martioli}}]{Adibekyan2021}
{Adibekyan}, V., {Dorn}, C., {Sousa}, S.~G., {et~al.} 2021, Science, 374, 330,
  \dodoi{10.1126/science.abg8794}

\bibitem[{{Adibekyan} {et~al.}(2012{\natexlab{a}}){Adibekyan}, {Sousa},
  {Santos}, {Delgado Mena}, {Gonz{\'a}lez Hern{\'a}ndez}, {Israelian}, {Mayor},
  \& {Khachatryan}}]{Adibekyan2012}
{Adibekyan}, V.~Z., {Sousa}, S.~G., {Santos}, N.~C., {et~al.}
  2012{\natexlab{a}}, \aap, 545, A32, \dodoi{10.1051/0004-6361/201219401}

\bibitem[{{Adibekyan} {et~al.}(2012{\natexlab{b}}){Adibekyan}, {Santos},
  {Sousa}, {Israelian}, {Delgado Mena}, {Gonz{\'a}lez Hern{\'a}ndez}, {Mayor},
  {Lovis}, \& {Udry}}]{Adibekyan2012b}
{Adibekyan}, V.~Z., {Santos}, N.~C., {Sousa}, S.~G., {et~al.}
  2012{\natexlab{b}}, \aap, 543, A89, \dodoi{10.1051/0004-6361/201219564}

\bibitem[{{Agnew} {et~al.}(2018){Agnew}, {Maddison}, \& {Horner}}]{Agnew2018}
{Agnew}, M.~T., {Maddison}, S.~T., \& {Horner}, J. 2018, \mnras, 477, 3646,
  \dodoi{10.1093/mnras/sty868}

\bibitem[{{Alvarez} \& {Plez}(1998)}]{Alvarez1998}
{Alvarez}, R., \& {Plez}, B. 1998, \aap, 330, 1109

\bibitem[{{Asplund}(2004)}]{Asplund2004}
{Asplund}, M. 2004, \memsai, 75, 300.
\newblock \doarXiv{astro-ph/0310444}

\bibitem[{{Astropy Collaboration} {et~al.}(2018){Astropy Collaboration},
  {Price-Whelan}, {Sip{\H o}cz}, {G{\"u}nther}, {Lim}, {Crawford}, {Conseil},
  {Shupe}, {Craig}, {Dencheva}, {Ginsburg}, {VanderPlas}, {Bradley},
  {P{\'e}rez-Su{\'a}rez}, {de Val-Borro}, {Aldcroft}, {Cruz}, {Robitaille},
  {Tollerud}, {Ardelean}, {Babej}, {Bach}, {Bachetti}, {Bakanov}, {Bamford},
  {Barentsen}, {Barmby}, {Baumbach}, {Berry}, {Biscani}, {Boquien}, {Bostroem},
  {Bouma}, {Brammer}, {Bray}, {Breytenbach}, {Buddelmeijer}, {Burke},
  {Calderone}, {Cano Rodr{\'{\i}}guez}, {Cara}, {Cardoso}, {Cheedella},
  {Copin}, {Corrales}, {Crichton}, {D'Avella}, {Deil}, {Depagne}, {Dietrich},
  {Donath}, {Droettboom}, {Earl}, {Erben}, {Fabbro}, {Ferreira}, {Finethy},
  {Fox}, {Garrison}, {Gibbons}, {Goldstein}, {Gommers}, {Greco}, {Greenfield},
  {Groener}, {Grollier}, {Hagen}, {Hirst}, {Homeier}, {Horton}, {Hosseinzadeh},
  {Hu}, {Hunkeler}, {Ivezi{\'c}}, {Jain}, {Jenness}, {Kanarek}, {Kendrew},
  {Kern}, {Kerzendorf}, {Khvalko}, {King}, {Kirkby}, {Kulkarni}, {Kumar},
  {Lee}, {Lenz}, {Littlefair}, {Ma}, {Macleod}, {Mastropietro}, {McCully},
  {Montagnac}, {Morris}, {Mueller}, {Mumford}, {Muna}, {Murphy}, {Nelson},
  {Nguyen}, {Ninan}, {N{\"o}the}, {Ogaz}, {Oh}, {Parejko}, {Parley}, {Pascual},
  {Patil}, {Patil}, {Plunkett}, {Prochaska}, {Rastogi}, {Reddy Janga},
  {Sabater}, {Sakurikar}, {Seifert}, {Sherbert}, {Sherwood-Taylor}, {Shih},
  {Sick}, {Silbiger}, {Singanamalla}, {Singer}, {Sladen}, {Sooley},
  {Sornarajah}, {Streicher}, {Teuben}, {Thomas}, {Tremblay}, {Turner},
  {Terr{\'o}n}, {van Kerkwijk}, {de la Vega}, {Watkins}, {Weaver}, {Whitmore},
  {Woillez}, {Zabalza}, \& {Astropy Contributors}}]{2018AJ....156..123A}
{Astropy Collaboration}, {Price-Whelan}, A.~M., {Sip{\H o}cz}, B.~M., {et~al.}
  2018, \aj, 156, 123, \dodoi{10.3847/1538-3881/aabc4f}

\bibitem[{{Bailer-Jones} {et~al.}(2021){Bailer-Jones}, {Rybizki}, {Fouesneau},
  {Demleitner}, \& {Andrae}}]{Bailer21}
{Bailer-Jones}, C.~A.~L., {Rybizki}, J., {Fouesneau}, M., {Demleitner}, M., \&
  {Andrae}, R. 2021, \aj, 161, 147, \dodoi{10.3847/1538-3881/abd806}

\bibitem[{{Bang} {et~al.}(2018){Bang}, {Lee}, {Jeong}, {Han}, \&
  {Park}}]{Bang18}
{Bang}, T.-Y., {Lee}, B.-C., {Jeong}, G.-h., {Han}, I., \& {Park}, M.-G. 2018,
  Journal of Korean Astronomical Society, 51, 17,
  \dodoi{10.5303/JKAS.2018.51.1.17}

\bibitem[{{Bashi} \& {Zucker}(2022)}]{Bashi2022}
{Bashi}, D., \& {Zucker}, S. 2022, \mnras, 510, 3449,
  \dodoi{10.1093/mnras/stab3596}

\bibitem[{{Bedell} {et~al.}(2018){Bedell}, {Bean}, {Mel{\'e}ndez}, {Spina},
  {Ram{\'\i}rez}, {Asplund}, {Alves-Brito}, {dos Santos}, {Dreizler}, {Yong},
  {Monroe}, \& {Casagrande}}]{Bedell2018}
{Bedell}, M., {Bean}, J.~L., {Mel{\'e}ndez}, J., {et~al.} 2018, \apj, 865, 68,
  \dodoi{10.3847/1538-4357/aad908}

\bibitem[{{Benatti} {et~al.}(2020){Benatti}, {Damasso}, {Desidera}, {Marzari},
  {Biazzo}, {Claudi}, {Di Mauro}, {Lanza}, {Pinamonti}, {Barbato}, {Malavolta},
  {Poretti}, {Sozzetti}, {Affer}, {Bignamini}, {Bonomo}, {Borsa}, {Brogi},
  {Bruno}, {Carleo}, {Cosentino}, {Covino}, {Frustagli}, {Giacobbe},
  {Gonzalez}, {Gratton}, {Harutyunyan}, {Knapic}, {Leto}, {Lodi}, {Maggio},
  {Maldonado}, {Mancini}, {Martinez Fiorenzano}, {Micela}, {Molinari},
  {Molinaro}, {Nardiello}, {Nascimbeni}, {Pagano}, {Pedani}, {Piotto},
  {Rainer}, \& {Scandariato}}]{Benatti20}
{Benatti}, S., {Damasso}, M., {Desidera}, S., {et~al.} 2020, \aap, 639, A50,
  \dodoi{10.1051/0004-6361/202037939}

\bibitem[{{Bensby} {et~al.}(2003){Bensby}, {Feltzing}, \&
  {Lundstr{\"o}m}}]{Bensby2003}
{Bensby}, T., {Feltzing}, S., \& {Lundstr{\"o}m}, I. 2003, \aap, 410, 527,
  \dodoi{10.1051/0004-6361:20031213}

\bibitem[{{Bensby} {et~al.}(2005){Bensby}, {Feltzing}, {Lundstr{\"o}m}, \&
  {Ilyin}}]{Bensby2005}
{Bensby}, T., {Feltzing}, S., {Lundstr{\"o}m}, I., \& {Ilyin}, I. 2005, \aap,
  433, 185, \dodoi{10.1051/0004-6361:20040332}

\bibitem[{{Bensby} {et~al.}(2014){Bensby}, {Feltzing}, \& {Oey}}]{Bensby2014}
{Bensby}, T., {Feltzing}, S., \& {Oey}, M.~S. 2014, \aap, 562, A71,
  \dodoi{10.1051/0004-6361/201322631}

\bibitem[{{Bergemann} {et~al.}(2021){Bergemann}, {Hoppe}, {Semenova},
  {Carlsson}, {Yakovleva}, {Voronov}, {Bautista}, {Nemer}, {Belyaev},
  {Leenaarts}, {Mashonkina}, {Reiners}, \& {Ellwarth}}]{Bergemann21}
{Bergemann}, M., {Hoppe}, R., {Semenova}, E., {et~al.} 2021, \mnras, 508, 2236,
  \dodoi{10.1093/mnras/stab2160}

\bibitem[{{Bitsch} \& {Battistini}(2020)}]{Bitsch20}
{Bitsch}, B., \& {Battistini}, C. 2020, \aap, 633, A10,
  \dodoi{10.1051/0004-6361/201936463}

\bibitem[{{Boisse} {et~al.}(2012){Boisse}, {Pepe}, {Perrier}, {Queloz},
  {Bonfils}, {Bouchy}, {Santos}, {Arnold}, {Beuzit}, {D{\'\i}az}, {Delfosse},
  {Eggenberger}, {Ehrenreich}, {Forveille}, {H{\'e}brard}, {Lagrange}, {Lovis},
  {Mayor}, {Moutou}, {Naef}, {Santerne}, {S{\'e}gransan}, {Sivan}, \&
  {Udry}}]{Boisse12}
{Boisse}, I., {Pepe}, F., {Perrier}, C., {et~al.} 2012, \aap, 545, A55,
  \dodoi{10.1051/0004-6361/201118419}

\bibitem[{{Bovy}(2015)}]{Bovy15}
{Bovy}, J. 2015, \apjs, 216, 29, \dodoi{10.1088/0067-0049/216/2/29}

\bibitem[{{Bovy} {et~al.}(2012){Bovy}, {Allende Prieto}, {Beers}, {Bizyaev},
  {da Costa}, {Cunha}, {Ebelke}, {Eisenstein}, {Frinchaboy}, {Garc{\'\i}a
  P{\'e}rez}, {Girardi}, {Hearty}, {Hogg}, {Holtzman}, {Maia}, {Majewski},
  {Malanushenko}, {Malanushenko}, {M{\'e}sz{\'a}ros}, {Nidever}, {O'Connell},
  {O'Donnell}, {Oravetz}, {Pan}, {Rocha-Pinto}, {Schiavon}, {Schneider},
  {Schultheis}, {Skrutskie}, {Smith}, {Weinberg}, {Wilson}, \&
  {Zasowski}}]{Bovy2012}
{Bovy}, J., {Allende Prieto}, C., {Beers}, T.~C., {et~al.} 2012, \apj, 759,
  131, \dodoi{10.1088/0004-637X/759/2/131}

\bibitem[{{Bowler} {et~al.}(2010){Bowler}, {Johnson}, {Marcy}, {Henry}, {Peek},
  {Fischer}, {Clubb}, {Liu}, {Reffert}, {Schwab}, \& {Lowe}}]{Bowler10}
{Bowler}, B.~P., {Johnson}, J.~A., {Marcy}, G.~W., {et~al.} 2010, \apj, 709,
  396, \dodoi{10.1088/0004-637X/709/1/396}

\bibitem[{{Bressan} {et~al.}(2012){Bressan}, {Marigo}, {Girardi}, {Salasnich},
  {Dal Cero}, {Rubele}, \& {Nanni}}]{Bressan2012}
{Bressan}, A., {Marigo}, P., {Girardi}, L., {et~al.} 2012, \mnras, 427, 127,
  \dodoi{10.1111/j.1365-2966.2012.21948.x}

\bibitem[{{Brooke} {et~al.}(2013){Brooke}, {Bernath}, {Schmidt}, \&
  {Bacskay}}]{Brooke2013}
{Brooke}, J.~S.~A., {Bernath}, P.~F., {Schmidt}, T.~W., \& {Bacskay}, G.~B.
  2013, \jqsrt, 124, 11, \dodoi{10.1016/j.jqsrt.2013.02.025}

\bibitem[{{Burt} {et~al.}(2021){Burt}, {Feng}, {Holden}, {Mamajek}, {Huang},
  {Rosenthal}, {Wang}, {Butler}, {Vogt}, {Laughlin}, {Henry}, {Teske}, {Wang},
  {Crane}, \& {Shectman}}]{Burt21}
{Burt}, J., {Feng}, F., {Holden}, B., {et~al.} 2021, \aj, 161, 10,
  \dodoi{10.3847/1538-3881/abc2d0}

\bibitem[{{Butler} {et~al.}(2006){Butler}, {Wright}, {Marcy}, {Fischer},
  {Vogt}, {Tinney}, {Jones}, {Carter}, {Johnson}, {McCarthy}, \&
  {Penny}}]{Butler06}
{Butler}, R.~P., {Wright}, J.~T., {Marcy}, G.~W., {et~al.} 2006, \apj, 646,
  505, \dodoi{10.1086/504701}

\bibitem[{{Clark} {et~al.}(2021){Clark}, {Clert{\'e}}, {Hinkel}, {Unterborn},
  {Wittenmyer}, {Horner}, {Wright}, {Carter}, {Morton}, {Spina}, {Asplund},
  {Buder}, {Bland-Hawthorn}, {Casey}, {De Silva}, {D'Orazi}, {Duong}, {Hayden},
  {Freeman}, {Kos}, {Lewis}, {Lin}, {Lind}, {Martell}, {Sharma}, {Simpson},
  {Zucker}, {Zwitter}, {Tinney}, {Ting (丁源森)}, {Nordlander}, \&
  {Amarsi}}]{Clark21}
{Clark}, J.~T., {Clert{\'e}}, M., {Hinkel}, N.~R., {et~al.} 2021, \mnras, 504,
  4968, \dodoi{10.1093/mnras/stab1052}

\bibitem[{{Clegg} {et~al.}(1981){Clegg}, {Lambert}, \& {Tomkin}}]{Clegg1981}
{Clegg}, R.~E.~S., {Lambert}, D.~L., \& {Tomkin}, J. 1981, \apj, 250, 262,
  \dodoi{10.1086/159371}

\bibitem[{{Courcol} {et~al.}(2015){Courcol}, {Bouchy}, {Pepe}, {Santerne},
  {Delfosse}, {Arnold}, {Astudillo-Defru}, {Boisse}, {Bonfils}, {Borgniet},
  {Bourrier}, {Cabrera}, {Deleuil}, {Demangeon}, {D{\'\i}az}, {Ehrenreich},
  {Forveille}, {H{\'e}brard}, {Lagrange}, {Montagnier}, {Moutou}, {Rey},
  {Santos}, {S{\'e}gransan}, {Udry}, \& {Wilson}}]{Courcol15}
{Courcol}, B., {Bouchy}, F., {Pepe}, F., {et~al.} 2015, \aap, 581, A38,
  \dodoi{10.1051/0004-6361/201526329}

\bibitem[{{Cowley} {et~al.}(2021){Cowley}, {Bord}, \& {Y{\"u}ce}}]{Cowley2021}
{Cowley}, C.~R., {Bord}, D.~J., \& {Y{\"u}ce}, K. 2021, \aj, 161, 142,
  \dodoi{10.3847/1538-3881/abdf5d}

\bibitem[{{Cridland} {et~al.}(2019){Cridland}, {Eistrup}, \& {van
  Dishoeck}}]{Cridland2019}
{Cridland}, A.~J., {Eistrup}, C., \& {van Dishoeck}, E.~F. 2019, \aap, 627,
  A127, \dodoi{10.1051/0004-6361/201834378}

\bibitem[{{Cutri} {et~al.}(2014){Cutri}, {Wright}, {Conrow}, {Fowler},
  {Eisenhardt}, {Grillmair}, {Kirkpatrick}, {Masci}, {McCallon}, {Wheelock},
  {Fajardo-Acosta}, {Yan}, {Benford}, {Harbut}, {Jarrett}, {Lake}, {Leisawitz},
  {Ressler}, {Stanford}, {Tsai}, {Liu}, {Helou}, {Mainzer}, {Gettngs},
  {Gonzalez}, {Hoffman}, {Marsh}, {Padgett}, {Skrutskie}, {Beck}, {Papin}, \&
  {Wittman}}]{Cutri14}
{Cutri}, R.~M., {Wright}, E.~L., {Conrow}, T., {et~al.} 2014, VizieR Online
  Data Catalog, II/328

\bibitem[{{da Silva} {et~al.}(2015){da Silva}, {Milone}, \&
  {Rocha-Pinto}}]{daSilva2015}
{da Silva}, R., {Milone}, A. d.~C., \& {Rocha-Pinto}, H.~J. 2015, \aap, 580,
  A24, \dodoi{10.1051/0004-6361/201525770}

\bibitem[{{Delgado Mena} {et~al.}(2021){Delgado Mena}, {Adibekyan}, {Santos},
  {Tsantaki}, {Gonz{\'a}lez Hern{\'a}ndez}, {Sousa}, \& {Bertr{\'a}n de
  Lis}}]{DelgadoMena2021}
{Delgado Mena}, E., {Adibekyan}, V., {Santos}, N.~C., {et~al.} 2021, \aap, 655,
  A99, \dodoi{10.1051/0004-6361/202141588}

\bibitem[{{Delgado Mena} {et~al.}(2010){Delgado Mena}, {Israelian},
  {Gonz{\'a}lez Hern{\'a}ndez}, {Bond}, {Santos}, {Udry}, \&
  {Mayor}}]{DelgadoMena2010}
{Delgado Mena}, E., {Israelian}, G., {Gonz{\'a}lez Hern{\'a}ndez}, J.~I.,
  {et~al.} 2010, \apj, 725, 2349, \dodoi{10.1088/0004-637X/725/2/2349}

\bibitem[{{D{\"o}llinger} {et~al.}(2009){D{\"o}llinger}, {Hatzes}, {Pasquini},
  {Guenther}, {Hartmann}, \& {Girardi}}]{Dollinger09}
{D{\"o}llinger}, M.~P., {Hatzes}, A.~P., {Pasquini}, L., {et~al.} 2009, \aap,
  499, 935, \dodoi{10.1051/0004-6361/200810837}

\bibitem[{{Dulick} {et~al.}(2003){Dulick}, {Bauschlicher}, {Burrows}, {Sharp},
  {Ram}, \& {Bernath}}]{Dulick2003}
{Dulick}, M., {Bauschlicher}, Jr., C.~W., {Burrows}, A., {et~al.} 2003, \apj,
  594, 651, \dodoi{10.1086/376791}

\bibitem[{{Endl} {et~al.}(2016){Endl}, {Brugamyer}, {Cochran}, {MacQueen},
  {Robertson}, {Meschiari}, {Ramirez}, {Shetrone}, {Gullikson}, {Johnson},
  {Wittenmyer}, {Horner}, {Ciardi}, {Horch}, {Simon}, {Howell}, {Everett},
  {Caldwell}, \& {Castanheira}}]{Endl16}
{Endl}, M., {Brugamyer}, E.~J., {Cochran}, W.~D., {et~al.} 2016, \apj, 818, 34,
  \dodoi{10.3847/0004-637X/818/1/34}

\bibitem[{{Fulton} {et~al.}(2015){Fulton}, {Weiss}, {Sinukoff}, {Isaacson},
  {Howard}, {Marcy}, {Henry}, {Holden}, \& {Kibrick}}]{Fulton15}
{Fulton}, B.~J., {Weiss}, L.~M., {Sinukoff}, E., {et~al.} 2015, \apj, 805, 175,
  \dodoi{10.1088/0004-637X/805/2/175}

\bibitem[{{Gaia Collaboration} {et~al.}(2016){Gaia Collaboration}, {Prusti},
  {de Bruijne}, {Brown}, {Vallenari}, {Babusiaux}, {Bailer-Jones}, {Bastian},
  {Biermann}, {Evans}, {Eyer}, {Jansen}, {Jordi}, {Klioner}, {Lammers},
  {Lindegren}, {Luri}, {Mignard}, {Milligan}, {Panem}, {Poinsignon},
  {Pourbaix}, {Randich}, {Sarri}, {Sartoretti}, {Siddiqui}, {Soubiran},
  {Valette}, {van Leeuwen}, {Walton}, {Aerts}, {Arenou}, {Cropper}, {Drimmel},
  {H{\o}g}, {Katz}, {Lattanzi}, {O'Mullane}, {Grebel}, {Holland}, {Huc},
  {Passot}, {Bramante}, {Cacciari}, {Casta{\~n}eda}, {Chaoul}, {Cheek}, {De
  Angeli}, {Fabricius}, {Guerra}, {Hern{\'a}ndez}, {Jean-Antoine-Piccolo},
  {Masana}, {Messineo}, {Mowlavi}, {Nienartowicz}, {Ord{\'o}{\~n}ez-Blanco},
  {Panuzzo}, {Portell}, {Richards}, {Riello}, {Seabroke}, {Tanga},
  {Th{\'e}venin}, {Torra}, {Els}, {Gracia-Abril}, {Comoretto},
  {Garcia-Reinaldos}, {Lock}, {Mercier}, {Altmann}, {Andrae}, {Astraatmadja},
  {Bellas-Velidis}, {Benson}, {Berthier}, {Blomme}, {Busso}, {Carry},
  {Cellino}, {Clementini}, {Cowell}, {Creevey}, {Cuypers}, {Davidson}, {De
  Ridder}, {de Torres}, {Delchambre}, {Dell'Oro}, {Ducourant}, {Fr{\'e}mat},
  {Garc{\'\i}a-Torres}, {Gosset}, {Halbwachs}, {Hambly}, {Harrison}, {Hauser},
  {Hestroffer}, {Hodgkin}, {Huckle}, {Hutton}, {Jasniewicz}, {Jordan},
  {Kontizas}, {Korn}, {Lanzafame}, {Manteiga}, {Moitinho}, {Muinonen},
  {Osinde}, {Pancino}, {Pauwels}, {Petit}, {Recio-Blanco}, {Robin}, {Sarro},
  {Siopis}, {Smith}, {Smith}, {Sozzetti}, {Thuillot}, {van Reeven}, {Viala},
  {Abbas}, {Abreu Aramburu}, {Accart}, {Aguado}, {Allan}, {Allasia},
  {Altavilla}, {{\'A}lvarez}, {Alves}, {Anderson}, {Andrei}, {Anglada Varela},
  {Antiche}, {Antoja}, {Ant{\'o}n}, {Arcay}, {Atzei}, {Ayache}, {Bach},
  {Baker}, {Balaguer-N{\'u}{\~n}ez}, {Barache}, {Barata}, {Barbier}, {Barblan},
  {Baroni}, {Barrado y Navascu{\'e}s}, {Barros}, {Barstow}, {Becciani},
  {Bellazzini}, {Bellei}, {Bello Garc{\'\i}a}, {Belokurov}, {Bendjoya},
  {Berihuete}, {Bianchi}, {Bienaym{\'e}}, {Billebaud}, {Blagorodnova},
  {Blanco-Cuaresma}, {Boch}, {Bombrun}, {Borrachero}, {Bouquillon}, {Bourda},
  {Bouy}, {Bragaglia}, {Breddels}, {Brouillet}, {Br{\"u}semeister},
  {Bucciarelli}, {Budnik}, {Burgess}, {Burgon}, {Burlacu}, {Busonero}, {Buzzi},
  {Caffau}, {Cambras}, {Campbell}, {Cancelliere}, {Cantat-Gaudin}, {Carlucci},
  {Carrasco}, {Castellani}, {Charlot}, {Charnas}, {Charvet}, {Chassat},
  {Chiavassa}, {Clotet}, {Cocozza}, {Collins}, {Collins}, {Costigan}, {Crifo},
  {Cross}, {Crosta}, {Crowley}, {Dafonte}, {Damerdji}, {Dapergolas}, {David},
  {David}, {De Cat}, {de Felice}, {de Laverny}, {De Luise}, {De March}, {de
  Martino}, {de Souza}, {Debosscher}, {del Pozo}, {Delbo}, {Delgado},
  {Delgado}, {di Marco}, {Di Matteo}, {Diakite}, {Distefano}, {Dolding}, {Dos
  Anjos}, {Drazinos}, {Dur{\'a}n}, {Dzigan}, {Ecale}, {Edvardsson}, {Enke},
  {Erdmann}, {Escolar}, {Espina}, {Evans}, {Eynard Bontemps}, {Fabre},
  {Fabrizio}, {Faigler}, {Falc{\~a}o}, {Farr{\`a}s Casas}, {Faye}, {Federici},
  {Fedorets}, {Fern{\'a}ndez-Hern{\'a}ndez}, {Fernique}, {Fienga}, {Figueras},
  {Filippi}, {Findeisen}, {Fonti}, {Fouesneau}, {Fraile}, {Fraser}, {Fuchs},
  {Furnell}, {Gai}, {Galleti}, {Galluccio}, {Garabato}, {Garc{\'\i}a-Sedano},
  {Gar{\'e}}, {Garofalo}, {Garralda}, {Gavras}, {Gerssen}, {Geyer}, {Gilmore},
  {Girona}, {Giuffrida}, {Gomes}, {Gonz{\'a}lez-Marcos},
  {Gonz{\'a}lez-N{\'u}{\~n}ez}, {Gonz{\'a}lez-Vidal}, {Granvik}, {Guerrier},
  {Guillout}, {Guiraud}, {G{\'u}rpide}, {Guti{\'e}rrez-S{\'a}nchez}, {Guy},
  {Haigron}, {Hatzidimitriou}, {Haywood}, {Heiter}, {Helmi}, {Hobbs},
  {Hofmann}, {Holl}, {Holland}, {Hunt}, {Hypki}, {Icardi}, {Irwin}, {Jevardat
  de Fombelle}, {Jofr{\'e}}, {Jonker}, {Jorissen}, {Julbe}, {Karampelas},
  {Kochoska}, {Kohley}, {Kolenberg}, {Kontizas}, {Koposov}, {Kordopatis},
  {Koubsky}, {Kowalczyk}, {Krone-Martins}, {Kudryashova}, {Kull}, {Bachchan},
  {Lacoste-Seris}, {Lanza}, {Lavigne}, {Le Poncin-Lafitte}, {Lebreton},
  {Lebzelter}, {Leccia}, {Leclerc}, {Lecoeur-Taibi}, {Lemaitre}, {Lenhardt},
  {Leroux}, {Liao}, {Licata}, {Lindstr{\o}m}, {Lister}, {Livanou}, {Lobel},
  {L{\"o}ffler}, {L{\'o}pez}, {Lopez-Lozano}, {Lorenz}, {Loureiro},
  {MacDonald}, {Magalh{\~a}es Fernandes}, {Managau}, {Mann}, {Mantelet},
  {Marchal}, {Marchant}, {Marconi}, {Marie}, {Marinoni}, {Marrese},
  {Marschalk{\'o}}, {Marshall}, {Mart{\'\i}n-Fleitas}, {Martino}, {Mary},
  {Matijevi{\v{c}}}, {Mazeh}, {McMillan}, {Messina}, {Mestre}, {Michalik},
  {Millar}, {Miranda}, {Molina}, {Molinaro}, {Molinaro}, {Moln{\'a}r},
  {Moniez}, {Montegriffo}, {Monteiro}, {Mor}, {Mora}, {Morbidelli}, {Morel},
  {Morgenthaler}, {Morley}, {Morris}, {Mulone}, {Muraveva}, {Musella},
  {Narbonne}, {Nelemans}, {Nicastro}, {Noval}, {Ord{\'e}novic},
  {Ordieres-Mer{\'e}}, {Osborne}, {Pagani}, {Pagano}, {Pailler}, {Palacin},
  {Palaversa}, {Parsons}, {Paulsen}, {Pecoraro}, {Pedrosa}, {Pentik{\"a}inen},
  {Pereira}, {Pichon}, {Piersimoni}, {Pineau}, {Plachy}, {Plum}, {Poujoulet},
  {Pr{\v{s}}a}, {Pulone}, {Ragaini}, {Rago}, {Rambaux}, {Ramos-Lerate},
  {Ranalli}, {Rauw}, {Read}, {Regibo}, {Renk}, {Reyl{\'e}}, {Ribeiro},
  {Rimoldini}, {Ripepi}, {Riva}, {Rixon}, {Roelens}, {Romero-G{\'o}mez},
  {Rowell}, {Royer}, {Rudolph}, {Ruiz-Dern}, {Sadowski}, {Sagrist{\`a}
  Sell{\'e}s}, {Sahlmann}, {Salgado}, {Salguero}, {Sarasso}, {Savietto},
  {Schnorhk}, {Schultheis}, {Sciacca}, {Segol}, {Segovia}, {Segransan},
  {Serpell}, {Shih}, {Smareglia}, {Smart}, {Smith}, {Solano}, {Solitro},
  {Sordo}, {Soria Nieto}, {Souchay}, {Spagna}, {Spoto}, {Stampa}, {Steele},
  {Steidelm{\"u}ller}, {Stephenson}, {Stoev}, {Suess}, {S{\"u}veges}, {Surdej},
  {Szabados}, {Szegedi-Elek}, {Tapiador}, {Taris}, {Tauran}, {Taylor},
  {Teixeira}, {Terrett}, {Tingley}, {Trager}, {Turon}, {Ulla}, {Utrilla},
  {Valentini}, {van Elteren}, {Van Hemelryck}, {van Leeuwen}, {Varadi},
  {Vecchiato}, {Veljanoski}, {Via}, {Vicente}, {Vogt}, {Voss}, {Votruba},
  {Voutsinas}, {Walmsley}, {Weiler}, {Weingrill}, {Werner}, {Wevers},
  {Whitehead}, {Wyrzykowski}, {Yoldas}, {{\v{Z}}erjal}, {Zucker}, {Zurbach},
  {Zwitter}, {Alecu}, {Allen}, {Allende Prieto}, {Amorim},
  {Anglada-Escud{\'e}}, {Arsenijevic}, {Azaz}, {Balm}, {Beck}, {Bernstein},
  {Bigot}, {Bijaoui}, {Blasco}, {Bonfigli}, {Bono}, {Boudreault}, {Bressan},
  {Brown}, {Brunet}, {Bunclark}, {Buonanno}, {Butkevich}, {Carret}, {Carrion},
  {Chemin}, {Ch{\'e}reau}, {Corcione}, {Darmigny}, {de Boer}, {de Teodoro}, {de
  Zeeuw}, {Delle Luche}, {Domingues}, {Dubath}, {Fodor}, {Fr{\'e}zouls},
  {Fries}, {Fustes}, {Fyfe}, {Gallardo}, {Gallegos}, {Gardiol}, {Gebran},
  {Gomboc}, {G{\'o}mez}, {Grux}, {Gueguen}, {Heyrovsky}, {Hoar}, {Iannicola},
  {Isasi Parache}, {Janotto}, {Joliet}, {Jonckheere}, {Keil}, {Kim},
  {Klagyivik}, {Klar}, {Knude}, {Kochukhov}, {Kolka}, {Kos}, {Kutka}, {Lainey},
  {LeBouquin}, {Liu}, {Loreggia}, {Makarov}, {Marseille}, {Martayan},
  {Martinez-Rubi}, {Massart}, {Meynadier}, {Mignot}, {Munari}, {Nguyen},
  {Nordlander}, {Ocvirk}, {O'Flaherty}, {Olias Sanz}, {Ortiz}, {Osorio},
  {Oszkiewicz}, {Ouzounis}, {Palmer}, {Park}, {Pasquato}, {Peltzer}, {Peralta},
  {P{\'e}turaud}, {Pieniluoma}, {Pigozzi}, {Poels}, {Prat}, {Prod'homme},
  {Raison}, {Rebordao}, {Risquez}, {Rocca-Volmerange}, {Rosen}, {Ruiz-Fuertes},
  {Russo}, {Sembay}, {Serraller Vizcaino}, {Short}, {Siebert}, {Silva},
  {Sinachopoulos}, {Slezak}, {Soffel}, {Sosnowska}, {Strai{\v{z}}ys}, {ter
  Linden}, {Terrell}, {Theil}, {Tiede}, {Troisi}, {Tsalmantza}, {Tur},
  {Vaccari}, {Vachier}, {Valles}, {Van Hamme}, {Veltz}, {Virtanen}, {Wallut},
  {Wichmann}, {Wilkinson}, {Ziaeepour}, \& {Zschocke}}]{Gaia16}
{Gaia Collaboration}, {Prusti}, T., {de Bruijne}, J.~H.~J., {et~al.} 2016,
  \aap, 595, A1, \dodoi{10.1051/0004-6361/201629272}

\bibitem[{{Gaia Collaboration} {et~al.}(2021){Gaia Collaboration}, {Brown},
  {Vallenari}, {Prusti}, {de Bruijne}, {Babusiaux}, {Biermann}, {Creevey},
  {Evans}, {Eyer}, {Hutton}, {Jansen}, {Jordi}, {Klioner}, {Lammers},
  {Lindegren}, {Luri}, {Mignard}, {Panem}, {Pourbaix}, {Randich}, {Sartoretti},
  {Soubiran}, {Walton}, {Arenou}, {Bailer-Jones}, {Bastian}, {Cropper},
  {Drimmel}, {Katz}, {Lattanzi}, {van Leeuwen}, {Bakker}, {Cacciari},
  {Casta{\~n}eda}, {De Angeli}, {Ducourant}, {Fabricius}, {Fouesneau},
  {Fr{\'e}mat}, {Guerra}, {Guerrier}, {Guiraud}, {Jean-Antoine Piccolo},
  {Masana}, {Messineo}, {Mowlavi}, {Nicolas}, {Nienartowicz}, {Pailler},
  {Panuzzo}, {Riclet}, {Roux}, {Seabroke}, {Sordo}, {Tanga}, {Th{\'e}venin},
  {Gracia-Abril}, {Portell}, {Teyssier}, {Altmann}, {Andrae}, {Bellas-Velidis},
  {Benson}, {Berthier}, {Blomme}, {Brugaletta}, {Burgess}, {Busso}, {Carry},
  {Cellino}, {Cheek}, {Clementini}, {Damerdji}, {Davidson}, {Delchambre},
  {Dell'Oro}, {Fern{\'a}ndez-Hern{\'a}ndez}, {Galluccio}, {Garc{\'\i}a-Lario},
  {Garcia-Reinaldos}, {Gonz{\'a}lez-N{\'u}{\~n}ez}, {Gosset}, {Haigron},
  {Halbwachs}, {Hambly}, {Harrison}, {Hatzidimitriou}, {Heiter},
  {Hern{\'a}ndez}, {Hestroffer}, {Hodgkin}, {Holl}, {Jan{\ss}en}, {Jevardat de
  Fombelle}, {Jordan}, {Krone-Martins}, {Lanzafame}, {L{\"o}ffler}, {Lorca},
  {Manteiga}, {Marchal}, {Marrese}, {Moitinho}, {Mora}, {Muinonen}, {Osborne},
  {Pancino}, {Pauwels}, {Petit}, {Recio-Blanco}, {Richards}, {Riello},
  {Rimoldini}, {Robin}, {Roegiers}, {Rybizki}, {Sarro}, {Siopis}, {Smith},
  {Sozzetti}, {Ulla}, {Utrilla}, {van Leeuwen}, {van Reeven}, {Abbas}, {Abreu
  Aramburu}, {Accart}, {Aerts}, {Aguado}, {Ajaj}, {Altavilla}, {{\'A}lvarez},
  {{\'A}lvarez Cid-Fuentes}, {Alves}, {Anderson}, {Anglada Varela}, {Antoja},
  {Audard}, {Baines}, {Baker}, {Balaguer-N{\'u}{\~n}ez}, {Balbinot}, {Balog},
  {Barache}, {Barbato}, {Barros}, {Barstow}, {Bartolom{\'e}}, {Bassilana},
  {Bauchet}, {Baudesson-Stella}, {Becciani}, {Bellazzini}, {Bernet}, {Bertone},
  {Bianchi}, {Blanco-Cuaresma}, {Boch}, {Bombrun}, {Bossini}, {Bouquillon},
  {Bragaglia}, {Bramante}, {Breedt}, {Bressan}, {Brouillet}, {Bucciarelli},
  {Burlacu}, {Busonero}, {Butkevich}, {Buzzi}, {Caffau}, {Cancelliere},
  {C{\'a}novas}, {Cantat-Gaudin}, {Carballo}, {Carlucci}, {Carnerero},
  {Carrasco}, {Casamiquela}, {Castellani}, {Castro-Ginard}, {Castro Sampol},
  {Chaoul}, {Charlot}, {Chemin}, {Chiavassa}, {Cioni}, {Comoretto}, {Cooper},
  {Cornez}, {Cowell}, {Crifo}, {Crosta}, {Crowley}, {Dafonte}, {Dapergolas},
  {David}, {David}, {de Laverny}, {De Luise}, {De March}, {De Ridder}, {de
  Souza}, {de Teodoro}, {de Torres}, {del Peloso}, {del Pozo}, {Delbo},
  {Delgado}, {Delgado}, {Delisle}, {Di Matteo}, {Diakite}, {Diener},
  {Distefano}, {Dolding}, {Eappachen}, {Edvardsson}, {Enke}, {Esquej}, {Fabre},
  {Fabrizio}, {Faigler}, {Fedorets}, {Fernique}, {Fienga}, {Figueras},
  {Fouron}, {Fragkoudi}, {Fraile}, {Franke}, {Gai}, {Garabato},
  {Garcia-Gutierrez}, {Garc{\'\i}a-Torres}, {Garofalo}, {Gavras}, {Gerlach},
  {Geyer}, {Giacobbe}, {Gilmore}, {Girona}, {Giuffrida}, {Gomel}, {Gomez},
  {Gonzalez-Santamaria}, {Gonz{\'a}lez-Vidal}, {Granvik},
  {Guti{\'e}rrez-S{\'a}nchez}, {Guy}, {Hauser}, {Haywood}, {Helmi}, {Hidalgo},
  {Hilger}, {H{\l}adczuk}, {Hobbs}, {Holland}, {Huckle}, {Jasniewicz},
  {Jonker}, {Juaristi Campillo}, {Julbe}, {Karbevska}, {Kervella}, {Khanna},
  {Kochoska}, {Kontizas}, {Kordopatis}, {Korn}, {Kostrzewa-Rutkowska},
  {Kruszy{\'n}ska}, {Lambert}, {Lanza}, {Lasne}, {Le Campion}, {Le Fustec},
  {Lebreton}, {Lebzelter}, {Leccia}, {Leclerc}, {Lecoeur-Taibi}, {Liao},
  {Licata}, {Lindstr{\o}m}, {Lister}, {Livanou}, {Lobel}, {Madrero Pardo},
  {Managau}, {Mann}, {Marchant}, {Marconi}, {Marcos Santos}, {Marinoni},
  {Marocco}, {Marshall}, {Martin Polo}, {Mart{\'\i}n-Fleitas}, {Masip},
  {Massari}, {Mastrobuono-Battisti}, {Mazeh}, {McMillan}, {Messina},
  {Michalik}, {Millar}, {Mints}, {Molina}, {Molinaro}, {Moln{\'a}r},
  {Montegriffo}, {Mor}, {Morbidelli}, {Morel}, {Morris}, {Mulone}, {Munoz},
  {Muraveva}, {Murphy}, {Musella}, {Noval}, {Ord{\'e}novic}, {Orr{\`u}},
  {Osinde}, {Pagani}, {Pagano}, {Palaversa}, {Palicio}, {Panahi}, {Pawlak},
  {Pe{\~n}alosa Esteller}, {Penttil{\"a}}, {Piersimoni}, {Pineau}, {Plachy},
  {Plum}, {Poggio}, {Poretti}, {Poujoulet}, {Pr{\v{s}}a}, {Pulone}, {Racero},
  {Ragaini}, {Rainer}, {Raiteri}, {Rambaux}, {Ramos}, {Ramos-Lerate}, {Re
  Fiorentin}, {Regibo}, {Reyl{\'e}}, {Ripepi}, {Riva}, {Rixon}, {Robichon},
  {Robin}, {Roelens}, {Rohrbasser}, {Romero-G{\'o}mez}, {Rowell}, {Royer},
  {Rybicki}, {Sadowski}, {Sagrist{\`a} Sell{\'e}s}, {Sahlmann}, {Salgado},
  {Salguero}, {Samaras}, {Sanchez Gimenez}, {Sanna}, {Santove{\~n}a},
  {Sarasso}, {Schultheis}, {Sciacca}, {Segol}, {Segovia}, {S{\'e}gransan},
  {Semeux}, {Shahaf}, {Siddiqui}, {Siebert}, {Siltala}, {Slezak}, {Smart},
  {Solano}, {Solitro}, {Souami}, {Souchay}, {Spagna}, {Spoto}, {Steele},
  {Steidelm{\"u}ller}, {Stephenson}, {S{\"u}veges}, {Szabados}, {Szegedi-Elek},
  {Taris}, {Tauran}, {Taylor}, {Teixeira}, {Thuillot}, {Tonello}, {Torra},
  {Torra}, {Turon}, {Unger}, {Vaillant}, {van Dillen}, {Vanel}, {Vecchiato},
  {Viala}, {Vicente}, {Voutsinas}, {Weiler}, {Wevers}, {Wyrzykowski}, {Yoldas},
  {Yvard}, {Zhao}, {Zorec}, {Zucker}, {Zurbach}, \& {Zwitter}}]{Gaia21}
{Gaia Collaboration}, {Brown}, A.~G.~A., {Vallenari}, A., {et~al.} 2021, \aap,
  649, A1, \dodoi{10.1051/0004-6361/202039657}

\bibitem[{{Galland} {et~al.}(2005){Galland}, {Lagrange}, {Udry}, {Chelli},
  {Pepe}, {Beuzit}, \& {Mayor}}]{Galland05}
{Galland}, F., {Lagrange}, A.~M., {Udry}, S., {et~al.} 2005, \aap, 444, L21,
  \dodoi{10.1051/0004-6361:200500176}

\bibitem[{{Garc{\'\i}a P{\'e}rez} {et~al.}(2016){Garc{\'\i}a P{\'e}rez},
  {Allende Prieto}, {Holtzman}, {Shetrone}, {M{\'e}sz{\'a}ros}, {Bizyaev},
  {Carrera}, {Cunha}, {Garc{\'\i}a-Hern{\'a}ndez}, {Johnson}, {Majewski},
  {Nidever}, {Schiavon}, {Shane}, {Smith}, {Sobeck}, {Troup}, {Zamora},
  {Weinberg}, {Bovy}, {Eisenstein}, {Feuillet}, {Frinchaboy}, {Hayden},
  {Hearty}, {Nguyen}, {O'Connell}, {Pinsonneault}, {Wilson}, \&
  {Zasowski}}]{GarciaPerez16}
{Garc{\'\i}a P{\'e}rez}, A.~E., {Allende Prieto}, C., {Holtzman}, J.~A.,
  {et~al.} 2016, \aj, 151, 144, \dodoi{10.3847/0004-6256/151/6/144}

\bibitem[{{Gonz{\'a}lez Hern{\'a}ndez} {et~al.}(2013){Gonz{\'a}lez
  Hern{\'a}ndez}, {Delgado-Mena}, {Sousa}, {Israelian}, {Santos}, {Adibekyan},
  \& {Udry}}]{GonzalezHernandez2013}
{Gonz{\'a}lez Hern{\'a}ndez}, J.~I., {Delgado-Mena}, E., {Sousa}, S.~G.,
  {et~al.} 2013, \aap, 552, A6, \dodoi{10.1051/0004-6361/201220165}

\bibitem[{{Grevesse} {et~al.}(2007){Grevesse}, {Asplund}, \&
  {Sauval}}]{Grevesse07}
{Grevesse}, N., {Asplund}, M., \& {Sauval}, A.~J. 2007, \ssr, 130, 105,
  \dodoi{10.1007/s11214-007-9173-7}

\bibitem[{{Gustafsson} {et~al.}(2008){Gustafsson}, {Edvardsson}, {Eriksson},
  {J{\o}rgensen}, {Nordlund}, \& {Plez}}]{Gustafsson2008}
{Gustafsson}, B., {Edvardsson}, B., {Eriksson}, K., {et~al.} 2008, \aap, 486,
  951, \dodoi{10.1051/0004-6361:200809724}

\bibitem[{{Gustafsson} {et~al.}(1999){Gustafsson}, {Karlsson}, {Olsson},
  {Edvardsson}, \& {Ryde}}]{Gustafsson1999}
{Gustafsson}, B., {Karlsson}, T., {Olsson}, E., {Edvardsson}, B., \& {Ryde}, N.
  1999, \aap, 342, 426.
\newblock \doarXiv{astro-ph/9811303}

\bibitem[{{Hara} {et~al.}(2020){Hara}, {Bouchy}, {Stalport}, {Boisse},
  {Rodrigues}, {Delisle}, {Santerne}, {Henry}, {Arnold}, {Astudillo-Defru},
  {Borgniet}, {Bonfils}, {Bourrier}, {Brugger}, {Courcol}, {Dalal}, {Deleuil},
  {Delfosse}, {Demangeon}, {D{\'\i}az}, {Dumusque}, {Forveille}, {H{\'e}brard},
  {Hobson}, {Kiefer}, {Lopez}, {Mignon}, {Mousis}, {Moutou}, {Pepe}, {Rey},
  {Santos}, {S{\'e}gransan}, {Udry}, \& {Wilson}}]{Hara20}
{Hara}, N.~C., {Bouchy}, F., {Stalport}, M., {et~al.} 2020, \aap, 636, L6,
  \dodoi{10.1051/0004-6361/201937254}

\bibitem[{{H{\'e}brard} {et~al.}(2016){H{\'e}brard}, {Arnold}, {Forveille},
  {Correia}, {Laskar}, {Bonfils}, {Boisse}, {D{\'\i}az}, {Hagelberg},
  {Sahlmann}, {Santos}, {Astudillo-Defru}, {Borgniet}, {Bouchy}, {Bourrier},
  {Courcol}, {Delfosse}, {Deleuil}, {Demangeon}, {Ehrenreich}, {Gregorio},
  {Jovanovic}, {Labrevoir}, {Lagrange}, {Lovis}, {Lozi}, {Moutou},
  {Montagnier}, {Pepe}, {Rey}, {Santerne}, {S{\'e}gransan}, {Udry},
  {Vanhuysse}, {Vigan}, \& {Wilson}}]{Hebrard16}
{H{\'e}brard}, G., {Arnold}, L., {Forveille}, T., {et~al.} 2016, \aap, 588,
  A145, \dodoi{10.1051/0004-6361/201527585}

\bibitem[{{Heiter} {et~al.}(2015){Heiter}, {Lind}, {Asplund}, {Barklem},
  {Bergemann}, {Magrini}, {Masseron}, {Mikolaitis}, {Pickering}, \&
  {Ruffoni}}]{Heiter2015}
{Heiter}, U., {Lind}, K., {Asplund}, M., {et~al.} 2015, \physscr, 90, 054010,
  \dodoi{10.1088/0031-8949/90/5/054010}

\bibitem[{{Hinkel} {et~al.}(2020){Hinkel}, {Hartnett}, \& {Young}}]{Hinkel20}
{Hinkel}, N.~R., {Hartnett}, H.~E., \& {Young}, P.~A. 2020, \apjl, 900, L38,
  \dodoi{10.3847/2041-8213/abb3cb}

\bibitem[{{Hinkel} \& {Unterborn}(2018)}]{Hinkel2018}
{Hinkel}, N.~R., \& {Unterborn}, C.~T. 2018, \apj, 853, 83,
  \dodoi{10.3847/1538-4357/aaa5b4}

\bibitem[{{J{\"o}nsson} {et~al.}(2020){J{\"o}nsson}, {Holtzman}, {Allende
  Prieto}, {Cunha}, {Garc{\'\i}a-Hern{\'a}ndez}, {Hasselquist}, {Masseron},
  {Osorio}, {Shetrone}, {Smith}, {Stringfellow}, {Bizyaev}, {Edvardsson},
  {Majewski}, {M{\'e}sz{\'a}ros}, {Souto}, {Zamora}, {Beaton}, {Bovy}, {Donor},
  {Pinsonneault}, {Poovelil}, \& {Sobeck}}]{Jonsson20}
{J{\"o}nsson}, H., {Holtzman}, J.~A., {Allende Prieto}, C., {et~al.} 2020, \aj,
  160, 120, \dodoi{10.3847/1538-3881/aba592}

\bibitem[{{Joshi}(2007)}]{Joshi07}
{Joshi}, Y.~C. 2007, \mnras, 378, 768, \dodoi{10.1111/j.1365-2966.2007.11831.x}

\bibitem[{{Jurgenson} {et~al.}(2016){Jurgenson}, {Fischer}, {McCracken},
  {Sawyer}, {Giguere}, {Szymkowiak}, {Santoro}, \& {Muller}}]{Jurgenson2016}
{Jurgenson}, C., {Fischer}, D., {McCracken}, T., {et~al.} 2016, Journal of
  Astronomical Instrumentation, 5, 1650003, \dodoi{10.1142/S2251171716500033}

\bibitem[{{Kervella} {et~al.}(2019){Kervella}, {Arenou}, {Mignard}, \&
  {Th{\'e}venin}}]{Kervella2019}
{Kervella}, P., {Arenou}, F., {Mignard}, F., \& {Th{\'e}venin}, F. 2019, \aap,
  623, A72, \dodoi{10.1051/0004-6361/201834371}

\bibitem[{{Kokaia} {et~al.}(2020){Kokaia}, {Davies}, \& {Mustill}}]{Kokaia2020}
{Kokaia}, G., {Davies}, M.~B., \& {Mustill}, A.~J. 2020, \mnras, 492, 352,
  \dodoi{10.1093/mnras/stz3408}

\bibitem[{{Kolecki} \& {Wang}(2021)}]{Kolecki2021}
{Kolecki}, J.~R., \& {Wang}, J. 2021, arXiv e-prints, arXiv:2112.02031.
\newblock \doarXiv{2112.02031}

\bibitem[{{Kurucz}(1993)}]{Kurucz1993}
{Kurucz}, R. 1993, Diatomic Molecular Data for Opacity Calculations.~Kurucz
  CD-ROM No.~15.~Cambridge, Mass.: Smithsonian Astrophysical Observatory,
  1993., 15

\bibitem[{{Lagarde} {et~al.}(2021){Lagarde}, {Reyl{\'e}}, {Chiappini}, {Mor},
  {Anders}, {Figueras}, {Miglio}, {Romero-G{\'o}mez}, {Antoja}, {Cabral},
  {Salomon}, {Robin}, {Bienaym{\'e}}, {Soubiran}, {Cornu}, \&
  {Montillaud}}]{Lagarde2021}
{Lagarde}, N., {Reyl{\'e}}, C., {Chiappini}, C., {et~al.} 2021, \aap, 654, A13,
  \dodoi{10.1051/0004-6361/202039982}

\bibitem[{{Lee} {et~al.}(2014){Lee}, {Han}, {Park}, {Mkrtichian}, {Hatzes}, \&
  {Kim}}]{Lee14}
{Lee}, B.~C., {Han}, I., {Park}, M.~G., {et~al.} 2014, \aap, 566, A67,
  \dodoi{10.1051/0004-6361/201322608}

\bibitem[{{Lindegren} {et~al.}(2021){Lindegren}, {Klioner}, {Hern{\'a}ndez},
  {Bombrun}, {Ramos-Lerate}, {Steidelm{\"u}ller}, {Bastian}, {Biermann}, {de
  Torres}, {Gerlach}, {Geyer}, {Hilger}, {Hobbs}, {Lammers}, {McMillan},
  {Stephenson}, {Casta{\~n}eda}, {Davidson}, {Fabricius}, {Gracia-Abril},
  {Portell}, {Rowell}, {Teyssier}, {Torra}, {Bartolom{\'e}}, {Clotet},
  {Garralda}, {Gonz{\'a}lez-Vidal}, {Torra}, {Abbas}, {Altmann}, {Anglada
  Varela}, {Balaguer-N{\'u}{\~n}ez}, {Balog}, {Barache}, {Becciani}, {Bernet},
  {Bertone}, {Bianchi}, {Bouquillon}, {Brown}, {Bucciarelli}, {Busonero},
  {Butkevich}, {Buzzi}, {Cancelliere}, {Carlucci}, {Charlot}, {Cioni},
  {Crosta}, {Crowley}, {del Peloso}, {del Pozo}, {Drimmel}, {Esquej}, {Fienga},
  {Fraile}, {Gai}, {Garcia-Reinaldos}, {Guerra}, {Hambly}, {Hauser},
  {Jan{\ss}en}, {Jordan}, {Kostrzewa-Rutkowska}, {Lattanzi}, {Liao}, {Licata},
  {Lister}, {L{\"o}ffler}, {Marchant}, {Masip}, {Mignard}, {Mints}, {Molina},
  {Mora}, {Morbidelli}, {Murphy}, {Pagani}, {Panuzzo}, {Pe{\~n}alosa Esteller},
  {Poggio}, {Re Fiorentin}, {Riva}, {Sagrist{\`a} Sell{\'e}s}, {Sanchez
  Gimenez}, {Sarasso}, {Sciacca}, {Siddiqui}, {Smart}, {Souami}, {Spagna},
  {Steele}, {Taris}, {Utrilla}, {van Reeven}, \& {Vecchiato}}]{Lindegren21}
{Lindegren}, L., {Klioner}, S.~A., {Hern{\'a}ndez}, J., {et~al.} 2021, \aap,
  649, A2, \dodoi{10.1051/0004-6361/202039709}

\bibitem[{{Liu} {et~al.}(2021){Liu}, {Bitsch}, {Asplund}, {Liu}, {Murphy},
  {Yong}, {Ting}, \& {Feltzing}}]{Liu2021}
{Liu}, F., {Bitsch}, B., {Asplund}, M., {et~al.} 2021, \mnras, 508, 1227,
  \dodoi{10.1093/mnras/stab2471}

\bibitem[{{Liu} {et~al.}(2020){Liu}, {Yong}, {Asplund}, {Wang}, {Spina},
  {Acu{\~n}a}, {Mel{\'e}ndez}, \& {Ram{\'\i}rez}}]{Liu2020}
{Liu}, F., {Yong}, D., {Asplund}, M., {et~al.} 2020, \mnras, 495, 3961,
  \dodoi{10.1093/mnras/staa1420}

\bibitem[{{Luhn} {et~al.}(2019){Luhn}, {Bastien}, {Wright}, {Johnson},
  {Howard}, \& {Isaacson}}]{Luhn19}
{Luhn}, J.~K., {Bastien}, F.~A., {Wright}, J.~T., {et~al.} 2019, \aj, 157, 149,
  \dodoi{10.3847/1538-3881/aaf5d0}

\bibitem[{{Madhusudhan}(2019)}]{Madhusudhan19}
{Madhusudhan}, N. 2019, \araa, 57, 617,
  \dodoi{10.1146/annurev-astro-081817-051846}

\bibitem[{{Madhusudhan}(2021)}]{Madhusudhan2021a}
---. 2021, {ExoFrontiers; Big questions in exoplanetary science},
  \dodoi{10.1088/2514-3433/abfa8f}

\bibitem[{{Masseron} {et~al.}(2014){Masseron}, {Plez}, {Van Eck}, {Colin},
  {Daoutidis}, {Godefroid}, {Coheur}, {Bernath}, {Jorissen}, \&
  {Christlieb}}]{Masseron2014}
{Masseron}, T., {Plez}, B., {Van Eck}, S., {et~al.} 2014, \aap, 571, A47,
  \dodoi{10.1051/0004-6361/201423956}

\bibitem[{{Mel{\'e}ndez} {et~al.}(2009){Mel{\'e}ndez}, {Asplund}, {Gustafsson},
  \& {Yong}}]{Melendez2009}
{Mel{\'e}ndez}, J., {Asplund}, M., {Gustafsson}, B., \& {Yong}, D. 2009, \apjl,
  704, L66, \dodoi{10.1088/0004-637X/704/1/L66}

\bibitem[{{Mikolaitis} {et~al.}(2018){Mikolaitis}, {Tautvai{\v{s}}ien{\.{e}}},
  {Drazdauskas}, {Minkevi{\v{c}}i{\={u}}t{\.{e}}}, {Klebonas}, {Bagdonas},
  {Pak{\v{s}}tien{\.{e}}}, \& {Janulis}}]{Mikolaitis2018}
{Mikolaitis}, {\v{S}}., {Tautvai{\v{s}}ien{\.{e}}}, G., {Drazdauskas}, A.,
  {et~al.} 2018, \pasp, 130, 074202, \dodoi{10.1088/1538-3873/aabfb6}

\bibitem[{{Mikolaitis} {et~al.}(2014){Mikolaitis}, {Hill}, {Recio-Blanco}, {de
  Laverny}, {Allende Prieto}, {Kordopatis}, {Tautvai{\v s}iene}, {Romano},
  {Gilmore}, {Randich}, {Feltzing}, {Micela}, {Vallenari}, {Alfaro}, {Bensby},
  {Bragaglia}, {Flaccomio}, {Lanzafame}, {Pancino}, {Smiljanic}, {Bergemann},
  {Carraro}, {Costado}, {Damiani}, {Hourihane}, {Jofr{\'e}}, {Lardo},
  {Magrini}, {Maiorca}, {Morbidelli}, {Sbordone}, {Sousa}, {Worley}, \&
  {Zaggia}}]{Mikolaitis2014}
{Mikolaitis}, {\v S}., {Hill}, V., {Recio-Blanco}, A., {et~al.} 2014, \aap,
  572, A33, \dodoi{10.1051/0004-6361/201424093}

\bibitem[{{Mikolaitis} {et~al.}(2019){Mikolaitis}, {Drazdauskas},
  {Minkevi{\v{c}}i{\={u}}t{\.{e}}}, {Stonkut{\.{e}}},
  {Tautvai{\v{s}}ien{\.{e}}}, {Klebonas}, {Bagdonas}, {Pak{\v{s}}tien{\.{e}}},
  \& {Janulis}}]{Mikolaitis2019}
{Mikolaitis}, {\v{S}}., {Drazdauskas}, A., {Minkevi{\v{c}}i{\={u}}t{\.{e}}},
  R., {et~al.} 2019, \aap, 628, A49, \dodoi{10.1051/0004-6361/201835004}

\bibitem[{{Mints} \& {Hekker}(2017)}]{Mints2017}
{Mints}, A., \& {Hekker}, S. 2017, \aap, 604, A108,
  \dodoi{10.1051/0004-6361/201630090}

\bibitem[{{Mints} \& {Hekker}(2018)}]{Mints18}
---. 2018, \aap, 618, A54, \dodoi{10.1051/0004-6361/201832739}

\bibitem[{{Mishenina} {et~al.}(2021){Mishenina}, {Basak}, {Adibekyan},
  {Soubiran}, \& {Kovtyukh}}]{Mishenina2021}
{Mishenina}, T., {Basak}, N., {Adibekyan}, V., {Soubiran}, C., \& {Kovtyukh},
  V. 2021, \mnras, 504, 4252, \dodoi{10.1093/mnras/stab1171}

\bibitem[{{Nibauer} {et~al.}(2021){Nibauer}, {Baxter}, {Jain}, {Van Saders},
  {Beaton}, \& {Teske}}]{Nibauer2021}
{Nibauer}, J., {Baxter}, E.~J., {Jain}, B., {et~al.} 2021, in American
  Astronomical Society Meeting Abstracts, Vol.~53, American Astronomical
  Society Meeting Abstracts, 332.04

\bibitem[{{Nordstr{\"o}m} {et~al.}(2004){Nordstr{\"o}m}, {Mayor}, {Andersen},
  {Holmberg}, {Pont}, {J{\o}rgensen}, {Olsen}, {Udry}, \&
  {Mowlavi}}]{Nordstrom2004}
{Nordstr{\"o}m}, B., {Mayor}, M., {Andersen}, J., {et~al.} 2004, \aap, 418,
  989, \dodoi{10.1051/0004-6361:20035959}

\bibitem[{{Pereira} {et~al.}(2009){Pereira}, {Kiselman}, \&
  {Asplund}}]{Pereira2009}
{Pereira}, T.~M.~D., {Kiselman}, D., \& {Asplund}, M. 2009, \aap, 507, 417,
  \dodoi{10.1051/0004-6361/200912829}

\bibitem[{{Ram} {et~al.}(2014){Ram}, {Brooke}, {Bernath}, {Sneden}, \&
  {Lucatello}}]{Ram2014}
{Ram}, R.~S., {Brooke}, J.~S.~A., {Bernath}, P.~F., {Sneden}, C., \&
  {Lucatello}, S. 2014, \apjs, 211, 5, \dodoi{10.1088/0067-0049/211/1/5}

\bibitem[{{Recio-Blanco} {et~al.}(2014){Recio-Blanco}, {de Laverny},
  {Kordopatis}, {Helmi}, {Hill}, {Gilmore}, {Wyse}, {Adibekyan}, {Randich},
  {Asplund}, {Feltzing}, {Jeffries}, {Micela}, {Vallenari}, {Alfaro}, {Allende
  Prieto}, {Bensby}, {Bragaglia}, {Flaccomio}, {Koposov}, {Korn}, {Lanzafame},
  {Pancino}, {Smiljanic}, {Jackson}, {Lewis}, {Magrini}, {Morbidelli},
  {Prisinzano}, {Sacco}, {Worley}, {Hourihane}, {Bergemann}, {Costado},
  {Heiter}, {Joffre}, {Lardo}, {Lind}, \& {Maiorca}}]{Recio2014}
{Recio-Blanco}, A., {de Laverny}, P., {Kordopatis}, G., {et~al.} 2014, \aap,
  567, A5, \dodoi{10.1051/0004-6361/201322944}

\bibitem[{{Ricker} {et~al.}(2015){Ricker}, {Winn}, {Vanderspek}, {Latham},
  {Bakos}, {Bean}, {Berta-Thompson}, {Brown}, {Buchhave}, {Butler}, {Butler},
  {Chaplin}, {Charbonneau}, {Christensen-Dalsgaard}, {Clampin}, {Deming},
  {Doty}, {De Lee}, {Dressing}, {Dunham}, {Endl}, {Fressin}, {Ge}, {Henning},
  {Holman}, {Howard}, {Ida}, {Jenkins}, {Jernigan}, {Johnson}, {Kaltenegger},
  {Kawai}, {Kjeldsen}, {Laughlin}, {Levine}, {Lin}, {Lissauer}, {MacQueen},
  {Marcy}, {McCullough}, {Morton}, {Narita}, {Paegert}, {Palle}, {Pepe},
  {Pepper}, {Quirrenbach}, {Rinehart}, {Sasselov}, {Sato}, {Seager},
  {Sozzetti}, {Stassun}, {Sullivan}, {Szentgyorgyi}, {Torres}, {Udry}, \&
  {Villasenor}}]{Ricker2015}
{Ricker}, G.~R., {Winn}, J.~N., {Vanderspek}, R., {et~al.} 2015, Journal of
  Astronomical Telescopes, Instruments, and Systems, 1, 014003,
  \dodoi{10.1117/1.JATIS.1.1.014003}

\bibitem[{{Robertson} {et~al.}(2012){Robertson}, {Endl}, {Cochran}, {MacQueen},
  {Wittenmyer}, {Horner}, {Brugamyer}, {Simon}, {Barnes}, \&
  {Caldwell}}]{Robertson12}
{Robertson}, P., {Endl}, M., {Cochran}, W.~D., {et~al.} 2012, \apj, 749, 39,
  \dodoi{10.1088/0004-637X/749/1/39}

\bibitem[{{Rosenthal} {et~al.}(2021){Rosenthal}, {Fulton}, {Hirsch},
  {Isaacson}, {Howard}, {Dedrick}, {Sherstyuk}, {Blunt}, {Petigura}, {Knutson},
  {Behmard}, {Chontos}, {Crepp}, {Crossfield}, {Dalba}, {Fischer}, {Henry},
  {Kane}, {Kosiarek}, {Marcy}, {Rubenzahl}, {Weiss}, \& {Wright}}]{Rosenthal21}
{Rosenthal}, L.~J., {Fulton}, B.~J., {Hirsch}, L.~A., {et~al.} 2021, \apjs,
  255, 8, \dodoi{10.3847/1538-4365/abe23c}

\bibitem[{{Saffe} {et~al.}(2019){Saffe}, {Jofr{\'e}}, {Miquelarena}, {Jaque
  Arancibia}, {Flores}, {L{\'o}pez}, \& {Collado}}]{Saffe2019}
{Saffe}, C., {Jofr{\'e}}, E., {Miquelarena}, P., {et~al.} 2019, \aap, 625, A39,
  \dodoi{10.1051/0004-6361/201935352}

\bibitem[{{Santos} {et~al.}(2017){Santos}, {Adibekyan}, {Figueira},
  {Andreasen}, {Barros}, {Delgado-Mena}, {Demangeon}, {Faria}, {Oshagh},
  {Sousa}, {Viana}, \& {Ferreira}}]{Santos2017a}
{Santos}, N.~C., {Adibekyan}, V., {Figueira}, P., {et~al.} 2017, \aap, 603,
  A30, \dodoi{10.1051/0004-6361/201730761}

\bibitem[{{Sato} {et~al.}(2013){Sato}, {Omiya}, {Harakawa}, {Liu}, {Izumiura},
  {Kambe}, {Takeda}, {Yoshida}, {Itoh}, {Ando}, {Kokubo}, \& {Ida}}]{Sato13}
{Sato}, B., {Omiya}, M., {Harakawa}, H., {et~al.} 2013, \pasj, 65, 85,
  \dodoi{10.1093/pasj/65.4.85}

\bibitem[{{Schneider} \& {Bitsch}(2021)}]{Schneider21}
{Schneider}, A.~D., \& {Bitsch}, B. 2021, \aap, 654, A71,
  \dodoi{10.1051/0004-6361/202039640}

\bibitem[{{Sch{\"o}nrich} {et~al.}(2010){Sch{\"o}nrich}, {Binney}, \&
  {Dehnen}}]{Schonrich10}
{Sch{\"o}nrich}, R., {Binney}, J., \& {Dehnen}, W. 2010, \mnras, 403, 1829,
  \dodoi{10.1111/j.1365-2966.2010.16253.x}

\bibitem[{{Schulze} {et~al.}(2021){Schulze}, {Wang}, {Johnson}, {Gaudi},
  {Unterborn}, \& {Panero}}]{Schulze2021}
{Schulze}, J.~G., {Wang}, J., {Johnson}, J.~A., {et~al.} 2021, PSJ, 2, 113,
  \dodoi{10.3847/PSJ/abcaa8}

\bibitem[{{Seabroke} {et~al.}(2021){Seabroke}, {Fabricius}, {Teyssier},
  {Sartoretti}, {Katz}, {Cropper}, {Antoja}, {Benson}, {Smith}, {Dolding},
  {Gosset}, {Panuzzo}, {Th{\'e}venin}, {Allende Prieto}, {Blomme}, {Guerrier},
  {Huckle}, {Jean-Antoine}, {Haigron}, {Marchal}, {Baker}, {Damerdji}, {David},
  {Fr{\'e}mat}, {Jan{\ss}en}, {Jasniewicz}, {Lobel}, {Samaras}, {Plum},
  {Soubiran}, {Vanel}, {Zwitter}, {Ajaj}, {Caffau}, {Chemin}, {Royer},
  {Brouillet}, {Crifo}, {Guy}, {Hambly}, {Leclerc}, {Mastrobuono-Battisti}, \&
  {Viala}}]{Seabroke21}
{Seabroke}, G.~M., {Fabricius}, C., {Teyssier}, D., {et~al.} 2021, \aap, 653,
  A160, \dodoi{10.1051/0004-6361/202141008}

\bibitem[{{Sharma} {et~al.}(2021){Sharma}, {Hayden}, \&
  {Bland-Hawthorn}}]{Sharma2021}
{Sharma}, S., {Hayden}, M.~R., \& {Bland-Hawthorn}, J. 2021, \mnras, 507, 5882,
  \dodoi{10.1093/mnras/stab2015}

\bibitem[{{Skrutskie} {et~al.}(2006){Skrutskie}, {Cutri}, {Stiening},
  {Weinberg}, {Schneider}, {Carpenter}, {Beichman}, {Capps}, {Chester},
  {Elias}, {Huchra}, {Liebert}, {Lonsdale}, {Monet}, {Price}, {Seitzer},
  {Jarrett}, {Kirkpatrick}, {Gizis}, {Howard}, {Evans}, {Fowler}, {Fullmer},
  {Hurt}, {Light}, {Kopan}, {Marsh}, {McCallon}, {Tam}, {Van Dyk}, \&
  {Wheelock}}]{Skrutskie2006}
{Skrutskie}, M.~F., {Cutri}, R.~M., {Stiening}, R., {et~al.} 2006, \aj, 131,
  1163, \dodoi{10.1086/498708}

\bibitem[{{Smiljanic} {et~al.}(2014){Smiljanic}, {Korn}, {Bergemann}, {Frasca},
  {Magrini}, {Masseron}, {Pancino}, {Ruchti}, {San Roman}, {Sbordone}, {Sousa},
  {Tabernero}, {Tautvai{\v s}ien{\.e}}, {Valentini}, {Weber}, {Worley},
  {Adibekyan}, {Allende Prieto}, {Barisevi{\v c}ius}, {Biazzo},
  {Blanco-Cuaresma}, {Bonifacio}, {Bragaglia}, {Caffau}, {Cantat-Gaudin},
  {Chorniy}, {de Laverny}, {Delgado-Mena}, {Donati}, {Duffau}, {Franciosini},
  {Friel}, {Geisler}, {Gonz{\'a}lez Hern{\'a}ndez}, {Gruyters}, {Guiglion},
  {Hansen}, {Heiter}, {Hill}, {Jacobson}, {Jofre}, {J{\"o}nsson}, {Lanzafame},
  {Lardo}, {Ludwig}, {Maiorca}, {Mikolaitis}, {Montes}, {Morel}, {Mucciarelli},
  {Mu{\~n}oz}, {Nordlander}, {Pasquini}, {Puzeras}, {Recio-Blanco}, {Ryde},
  {Sacco}, {Santos}, {Serenelli}, {Sordo}, {Soubiran}, {Spina}, {Steffen},
  {Vallenari}, {Van Eck}, {Villanova}, {Gilmore}, {Randich}, {Asplund},
  {Binney}, {Drew}, {Feltzing}, {Ferguson}, {Jeffries}, {Micela}, {Negueruela},
  {Prusti}, {Rix}, {Alfaro}, {Babusiaux}, {Bensby}, {Blomme}, {Flaccomio},
  {Fran{\c c}ois}, {Irwin}, {Koposov}, {Walton}, {Bayo}, {Carraro}, {Costado},
  {Damiani}, {Edvardsson}, {Hourihane}, {Jackson}, {Lewis}, {Lind}, {Marconi},
  {Martayan}, {Monaco}, {Morbidelli}, {Prisinzano}, \&
  {Zaggia}}]{Smiljanic2014}
{Smiljanic}, R., {Korn}, A.~J., {Bergemann}, M., {et~al.} 2014, \aap, 570,
  A122, \dodoi{10.1051/0004-6361/201423937}

\bibitem[{{Sneden} {et~al.}(2014){Sneden}, {Lucatello}, {Ram}, {Brooke}, \&
  {Bernath}}]{Sneden2014}
{Sneden}, C., {Lucatello}, S., {Ram}, R.~S., {Brooke}, J.~S.~A., \& {Bernath},
  P. 2014, \apjs, 214, 26, \dodoi{10.1088/0067-0049/214/2/26}

\bibitem[{{Sneden}(1973)}]{Sneden1973}
{Sneden}, C.~A. 1973, PhD thesis, THE UNIVERSITY OF TEXAS AT AUSTIN.

\bibitem[{{Soubiran} {et~al.}(2016){Soubiran}, {Le Campion}, {Brouillet}, \&
  {Chemin}}]{Soubiran2016}
{Soubiran}, C., {Le Campion}, J.-F., {Brouillet}, N., \& {Chemin}, L. 2016,
  \aap, 591, A118, \dodoi{10.1051/0004-6361/201628497}

\bibitem[{{Stassun} {et~al.}(2017){Stassun}, {Collins}, \& {Gaudi}}]{Stassun17}
{Stassun}, K.~G., {Collins}, K.~A., \& {Gaudi}, B.~S. 2017, \aj, 153, 136,
  \dodoi{10.3847/1538-3881/aa5df3}

\bibitem[{{Stetson} \& {Pancino}(2008)}]{Stetson2008}
{Stetson}, P.~B., \& {Pancino}, E. 2008, \pasp, 120, 1332,
  \dodoi{10.1086/596126}

\bibitem[{{Stonkut{\.{e}}} {et~al.}(2020){Stonkut{\.{e}}}, {Chorniy},
  {Tautvai{\v{s}}ien{\.{e}}}, {Drazdauskas}, {Minkevi{\v{c}}i{\={u}}t{\.{e}}},
  {Mikolaitis}, {Kjeldsen}, {Essen}, {Pak{\v{s}}tien{\.{e}}}, \&
  {Bagdonas}}]{Stonkute2020}
{Stonkut{\.{e}}}, E., {Chorniy}, Y., {Tautvai{\v{s}}ien{\.{e}}}, G., {et~al.}
  2020, \aj, 159, 90, \dodoi{10.3847/1538-3881/ab6a19}

\bibitem[{{Su{\'a}rez-Andr{\'e}s} {et~al.}(2016){Su{\'a}rez-Andr{\'e}s},
  {Israelian}, {Gonz{\'a}lez Hern{\'a}ndez}, {Adibekyan}, {Delgado Mena},
  {Santos}, \& {Sousa}}]{Suarez-Andres2016}
{Su{\'a}rez-Andr{\'e}s}, L., {Israelian}, G., {Gonz{\'a}lez Hern{\'a}ndez},
  J.~I., {et~al.} 2016, \aap, 591, A69, \dodoi{10.1051/0004-6361/201628455}

\bibitem[{{Su{\'a}rez-Andr{\'e}s} {et~al.}(2017){Su{\'a}rez-Andr{\'e}s},
  {Israelian}, {Gonz{\'a}lez Hern{\'a}ndez}, {Adibekyan}, {Delgado Mena},
  {Santos}, \& {Sousa}}]{Suarez2017}
---. 2017, \aap, 599, A96, \dodoi{10.1051/0004-6361/201629434}

\bibitem[{{Su{\'a}rez-Andr{\'e}s} {et~al.}(2018){Su{\'a}rez-Andr{\'e}s},
  {Israelian}, {Gonz{\'a}lez Hern{\'a}ndez}, {Adibekyan}, {Delgado Mena},
  {Santos}, \& {Sousa}}]{Suarez18}
---. 2018, \aap, 614, A84, \dodoi{10.1051/0004-6361/201730743}

\bibitem[{{Tautvai{\v{s}}ien{\.{e}}} {et~al.}(2010){Tautvai{\v{s}}ien{\.{e}}},
  {Edvardsson}, {Puzeras}, {Barisevi{\v{c}}ius}, \& {Ilyin}}]{Tautvaisiene2010}
{Tautvai{\v{s}}ien{\.{e}}}, G., {Edvardsson}, B., {Puzeras}, E.,
  {Barisevi{\v{c}}ius}, G., \& {Ilyin}, I. 2010, \mnras, 409, 1213,
  \dodoi{10.1111/j.1365-2966.2010.17381.x}

\bibitem[{{Tautvai{\v{s}}ien{\.{e}}} {et~al.}(2020){Tautvai{\v{s}}ien{\.{e}}},
  {Mikolaitis}, {Drazdauskas}, {Stonkut{\.{e}}},
  {Minkevi{\v{c}}i{\={u}}t{\.{e}}}, {Kjeldsen}, {Brogaard}, {von Essen},
  {Grundahl}, {Pak{\v{s}}tien{\.{e}}}, {Bagdonas}, \&
  {V{\'a}zquez}}]{Tautvaisiene20}
{Tautvai{\v{s}}ien{\.{e}}}, G., {Mikolaitis}, {\v{S}}., {Drazdauskas}, A.,
  {et~al.} 2020, \apjs, 248, 19, \dodoi{10.3847/1538-4365/ab8b67}

\bibitem[{{Thiabaud} {et~al.}(2015){Thiabaud}, {Marboeuf}, {Alibert}, {Leya},
  \& {Mezger}}]{Thiabaud15}
{Thiabaud}, A., {Marboeuf}, U., {Alibert}, Y., {Leya}, I., \& {Mezger}, K.
  2015, \aap, 580, A30, \dodoi{10.1051/0004-6361/201525963}

\bibitem[{{Tucci Maia} {et~al.}(2014){Tucci Maia}, {Mel{\'e}ndez}, \&
  {Ram{\'\i}rez}}]{TucciMaia2014}
{Tucci Maia}, M., {Mel{\'e}ndez}, J., \& {Ram{\'\i}rez}, I. 2014, \apjl, 790,
  L25, \dodoi{10.1088/2041-8205/790/2/L25}

\bibitem[{{Turrini} {et~al.}(2021){Turrini}, {Schisano}, {Fonte}, {Molinari},
  {Politi}, {Fedele}, {Pani{\'c}}, {Kama}, {Changeat}, \&
  {Tinetti}}]{Turrini2021}
{Turrini}, D., {Schisano}, E., {Fonte}, S., {et~al.} 2021, \apj, 909, 40,
  \dodoi{10.3847/1538-4357/abd6e5}

\bibitem[{{Unterborn} \& {Panero}(2019)}]{Unterborn19}
{Unterborn}, C.~T., \& {Panero}, W.~R. 2019, Journal of Geophysical Research
  (Planets), 124, 1704, \dodoi{10.1029/2018JE005844}

\bibitem[{{Wenger} {et~al.}(2000){Wenger}, {Ochsenbein}, {Egret}, {Dubois},
  {Bonnarel}, {Borde}, {Genova}, {Jasniewicz}, {Lalo{\"e}}, {Lesteven}, \&
  {Monier}}]{Wenger2000}
{Wenger}, M., {Ochsenbein}, F., {Egret}, D., {et~al.} 2000, \aaps, 143, 9,
  \dodoi{10.1051/aas:2000332}

\bibitem[{{Zhang} {et~al.}(2021){Zhang}, {Chen}, {Zhao}, {Zhao}, {Liang}, {Li},
  {Wu}, {Luo}, \& {Wang}}]{Zhang2021}
{Zhang}, H.-P., {Chen}, Y.-Q., {Zhao}, G., {et~al.} 2021, Research in Astronomy
  and Astrophysics, 21, 153, \dodoi{10.1088/1674-4527/21/6/153}

\end{thebibliography}
\bibliographystyle{aasjournal}

\end{document}